\documentclass[11pt,oneside,a4paper]{article}

\usepackage[margin=2.5cm]{geometry}

\usepackage{setspace}
\usepackage{soul}
\usepackage{amsfonts}
\usepackage{dsfont}
\usepackage[cmex10]{amsmath}
\usepackage{amssymb,amsbsy,mathrsfs}

\usepackage{hyperref}

\usepackage{fancyhdr}

\usepackage{comment}

\usepackage{booktabs}  
\usepackage{array}     

\usepackage{mathptmx}
\usepackage{slashbox}
\usepackage{multirow}
\usepackage[table]{xcolor}
\usepackage{colortbl}

\usepackage{algorithm}
\usepackage{algpseudocode}
\algnewcommand{\algorithmicgoto}{\textbf{go to}}%
\algnewcommand{\Goto}[1]{\algorithmicgoto~\ref{#1}}%

\newtheorem{definition}{Definition}

\usepackage{graphicx}
\graphicspath{ {./} }
\DeclareGraphicsExtensions{.pdf}

\begin{document}

\title{Optimizing the Gossip Algorithm with Non-Uniform Clock Distribution over Classical \& Quantum Networks}

\author{Saber Jafarizadeh  \\ {saber.jafarizadeh@sydney.edu.au} }
\date{}%

\markboth{S. Jafarizadeh}{Gossip Algorithm with Non-Uniform Clock Distribution}

\providecommand{\keywords}[1]{\textbf{\textit{Index terms---}} #1}

\maketitle

\bibliographystyle{ieeetran}

\begin{abstract}

Distributed gossip algorithm has been studied in literature for practical implementation of the distributed consensus algorithm as a fundamental algorithm for the purpose of in-network collaborative processing.
This paper focuses on optimizing the convergence rate of the gossip algorithm for both classical and quantum networks.
A novel model of the gossip algorithm with non-uniform clock distribution is proposed which can reach the optimal convergence rate of the continuous-time consensus algorithm.
It is described that how the non-uniform clock distribution is achievable by modifying the rate of the Poisson process modeling the clock of the gossip algorithm.
The minimization problem for optimizing the asymptotic convergence rate of the proposed gossip algorithm and its corresponding semidefinite programming formulation is addressed analytically.
It is shown that the optimal results obtained for uniform clock distribution are suboptimal compared to those of the non-uniform one and for non-uniform distribution the optimal answer is not unique i.e. there is more than one set of probabilities that can achieve the optimal convergence rate.
Based on the optimal continuous-time consensus algorithm and the detailed balance property, an effective method of obtaining one of these optimal answers is proposed.
Regarding quantum gossip algorithm, by expanding the density matrix in terms of the generalized Gell-Mann matrices, the evolution equation of the quantum gossip algorithm is transformed to the state update equation of the classical gossip algorithm.
By defining the quantum gossip operator, the original optimization problem is formulated as a convex optimization problem, which can be addressed by semidefinite programming.

\end{abstract}


\section{Introduction}

Many manmade systems of the present day are networks of a large number of cooperating standalone units.
Examples of such systems include power grids, transportation and water distribution networks, telephone networks, the Internet, the global financial network and social networks \cite{KocarevBookComplexConsensus2013,WeiBookVehicleConsensus2008}.
The appealing features of networked systems are their natural flexibility, reliability and robustness to failure, easy re-configuration, and low cost.
Distributed consensus algorithm \cite{TsitsiklisThesis1984,TsitsiklisAutomaticControl1986} is one of the fundamental algorithms for the purpose of in-network collaborative processing in the context of coordinated control of networked systems modeled as networks of autonomous agents.
Using distributed consensus algorithm, agents achieve a stable and reliable opinion regarding a phenomenon, through the local exchange of information with their neighbors.
Distributed consensus  algorithm has found applications in numerous fields including vehicle formation \cite{Gossip22Application2004}, multiagent collaboration and flocking \cite{Gossip20Application2003,Gossip21Application2004}, data fusion and tracking \cite{Gossip23Application2006}, distributed inference \cite{Gossip24Application2008} and load balancing \cite{Gossip19Application1989} .

However, there are major challenges in the implementation of distributed consensus algorithm to enable agents to exchange their state with all their neighbors in every round of the algorithm.
Distributed gossip algorithm \cite{Boyd06Gossip} emerges as a mean for implementing the distributed consensus algorithm.
There is an extensive body of literature on consensus and gossip algorithms, see the surveys \cite{RenGossipSurvey2007,OlfatiSaberGossipSurvay2007,CaoGossipSurvey2013,DimakisGossipSurvey2010}, and references therein.

One of the pioneering works on the gossip algorithms is \cite{BoydInfocom2005} where the authors provide the bound on the running time of the gossip algorithm to converge to the global average with certain accuracy.
\cite{Carli2010Automatica}  analyzed the effects of quantized communication on the gossip algorithm.
\cite{DimakisGeographic2008} enhanced the convergence rate of the gossip algorithm by exploiting the geographic information of the network.
In \cite{AndersonAutomatica2013} accelerated gossip algorithms are proposed where local memory is exploited by installing shift-registers at each agent.
\cite{DimakisMobility2012} studied the effect of node mobility on the convergence rate of the gossip algorithm.

Gossip algorithm has been employed for different applications.
\cite{DuchiAutomaticControl2012} developed several distributed gossip-based algorithms to solve convex optimization problems defined over networks.
\cite{KarSignalprocessing2011} introduced a gossip-based algorithm performing distributed Kalman filtering for networked systems.
\cite{KarDistributedEstimation2011IEEESignalProcessing} employed gossip algorithm for distributed estimation in static distributed random fields.
\cite{AndersonDeterministic2011} studied different deterministic gossiping protocols for avoiding deadlocks and achieving consensus under different assumptions.

In the conventional gossip algorithm \cite{Boyd06Gossip}, each agent updates its state with one of its neighbors in random when its clock ticks. The clock ticks are at the times of a Poisson process.
The probability that each agent selects one of its neighbors is referred to as the transition probabilities and the probability that each agent starts the state updating process is the clock distribution probabilities.
In the gossip algorithm introduced in \cite{Boyd06Gossip}, the clocks of agents follow the Poisson processes with the same rate.
This results in uniform clock distribution, which in turn leaves the transition probabilities as the only optimization parameters.
The main motivation behind the analysis presented in this paper is to propose and optimize a gossip algorithm that can achieve the convergence rate of the optimal continuous-time consensus algorithm over the same network topology.
To do so, in the gossip algorithm proposed in this paper, the clock distribution probabilities are considered as variables in the optimization problem.
This results in a more general framework for optimizing the gossip algorithm where the case of uniform clock distribution \cite{Boyd06Gossip} is a subset of this problem.

First we introduce the gossip algorithm with non-uniform clock distribution over classical networks.
Then we explain how the Poisson process is the most suitable for the clock tick model of the gossip algorithm, and subsequently how non-uniform clock distribution can be achieved by modifying the rate of the Poison process according to the clock distribution probabilities.
After formulating the asymptotic convergence rate of the gossip algorithm, the minimization problem and its corresponding semidefinite programming problem for optimizing the asymptotic convergence rate of the gossip algorithm over classical networks is presented.

We have provided the closed-form formulas for the probabilities and the convergence rate of the gossip algorithm for both cases of uniform and non-uniform clock distributions over different network topologies.
This is done to show that the optimal results obtained for uniform clock distribution are suboptimal compared to those of the non-uniform clock distribution.
For the case of non-uniform clock distribution, the optimal answer is not necessarily unique i.e. there is more than one set of transition and clock distribution probabilities that can achieve the fastest convergence rate.
This is due to the fact that the number of variables in the semidefinite programming formulation of the problem is more than the number of equations.
We have shown that for gossip algorithm with non-uniform clock distribution, one of the optimal answers is achievable by using the optimal results of the continuous-time consensus algorithm \cite{SaberQConsensusContinuous} and ensuring that the transition and clock distribution probabilities satisfy the detailed balance property.

The concept of consensus and gossip algorithms over quantum networks is a relatively new topic.
The initial activities in this field have been initiated by the recent progress in the field of quantum information science and quantum distributed computing \cite{Broadbent2008,Buhrman2003,Denchev:2008}.
This development is driven by the fact that centralized and big quantum computers with numerous qubits are expensive and difficult to build and maintain.
An alternative to such systems is a network of smaller quantum computers which is analyzed as networks of Quantum nodes.
Consensus and gossip problems are brought to the quantum domain in \cite{MazzarellaCDC2013,PetersenRef15,MazzarellaArXiv} where they are reinterpreted as a symmetrization problem.
Authors in \cite{MazzarellaCDC2013,PetersenRef15,MazzarellaArXiv} derive the general conditions for convergence of the consensus algorithm for quantum networks.
Authors in \cite{Petersen2015IEEETranAutControl,Petersen2015ACCPartI,Petersen2015ACCPartII} took a different approach and employed the induced graphs of the quantum interaction graph.
By doing so, they could relate the consensus problem over a network of N-qubit to its equivalent classical consensus problem, % and formulated the convergence rate of the resultant classical problem as a convex optimization problem.
and established the necessary and sufficient conditions for the exponential and asymptotic convergence rate of the quantum consensus.

In \cite{SaberQConsensusContinuous}, the convergence rate of the continuous-time quantum consensus is optimized, and it is shown that the optimal convergence rate is independent of the value of $d$ in qudits.
They have shown that the induced graphs are the Schreier graphs and using the Young Tabloids; they have categorized all possible induced graphs.
Furthermore, by extending the proof of the Aldous' conjecture \cite{ProofAldous} to all possible induced graphs, they have reduced the optimization of the continuous-time quantum consensus to optimizing the second smallest eigenvalue of the Laplacian of the induced graph corresponding to partition $(N-1, 1)$.
For a wide range of topologies, they have analytically addressed the semidefinite programming formulation of the reduced optimization problem.
\cite{SaberQConsensusDiscrete} studies the optimization of the discrete-time model of the quantum consensus over a quantum network with $N$ qudits.
In contrast to the results obtained for the continuous time model \cite{SaberQConsensusContinuous}, in \cite{SaberQConsensusDiscrete} it is shown that the convergence rate of the consensus algorithm depends on the value of $d$ is qudits.
Exploiting the Specht module representation of partitions of $N$, in \cite{SaberQConsensusDiscrete} it is shown that the Laplacian matrix corresponding to partition $(1, 1, \ldots, 1)$ includes the corresponding spectrum of all other partitions.
Based on this result and the generalization of the Aldous' conjecture to all partitions of integer $N$, % they have generalized the Aldous' conjecture to all partitions of integer $N$ and
they have shown that the problem of optimizing the convergence rate of the discrete-time quantum consensus reduces to optimizing the Second Largest Eigenvalue Modulus (SLEM) of the weight matrix.
By providing the semidefinite programming formulation of the resultant problem, they have addressed it analytically.
The analysis presented in this paper aim at optimizing the convergence rate of the quantum gossip algorithm to that of its quantum consensus state.
By expanding the density matrix in terms of the generalized Gell-Mann matrices, we transform the evolution equation of the quantum gossip algorithm to the state update equation of the classical gossip algorithm.
After defining the quantum gossip operator, we have formulated the original optimization problem as a convex optimization problem. % which depending on the underlying graph of the quantum network can be solved using semidefinite programming or linear programming.
We have shown that for different topologies, the resultant convex optimization problem can be solved using semidefinite programming and linear programming.

In section \ref{sec:ClassicalGossip} our model for the gossip algorithm with non-uniform clock distribution is introduced.
Followed by derivation of the optimization problem for optimizing the convergence rate of the gossip algorithm and the semidefinite programming formulation of the obtained optimization problem.
Section \ref{sec:QuantumGossipAlgorithm} presents the quantum gossip algorithm and explains how it can be modeled as a classical gossip algorithm.
Formulation of the convex optimization problem for optimizing the convergence rate of the quantum gossip algorithm is also presented in this section.
In section \ref{sec:OptimizationSectionComplete} analytical optimization of the gossip algorithm along with closed-form expressions for the optimal probabilities and convergence rate for a range of topologies have been presented.
Section \ref{sec:Conclusion} concludes the paper.

\section{Classical Gossip Algorithm (CGA)}
\label{sec:ClassicalGossip}

In our model for the gossip algorithm, once the clock in agent $i$ ticks, it selects one of its neighbors, (agent $j$) at random with probability $P_{i,j}$ and communicates with the selected neighboring agent.
Upon communication, two agents update their states $(\boldsymbol{x}_i(k), \boldsymbol{x}_j(k))$ to the average of their current values $(\boldsymbol{x}_i(k) + \boldsymbol{x}_j(k))/2$.
This can be described by the following equation,
\begin{equation}
    \begin{aligned}
        \label{eq:Gossip1}
        \boldsymbol{x}(k+1) = \boldsymbol{W}_{i,j} \times \boldsymbol{x}(k),
    \end{aligned}
\end{equation}
where
\begin{equation}
    \begin{aligned}
        \label{eq:Gossip2}
        \boldsymbol{W}_{i,j} = \boldsymbol{I} - \left[(\boldsymbol{e}_i - \boldsymbol{e}_j)\times(\boldsymbol{e}_i - \boldsymbol{e}_j)^T\right]/2.
    \end{aligned}
\end{equation}
$\boldsymbol{I}$ is the identity matrix and $\boldsymbol{e}_i$ is a column vector where its $i$-th element is $1$ and the rest is zero.
We refer to each one of the matrices $\boldsymbol{W}_{i,j}$ as an averaging matrix.
Defining $\Pi_{ij}$ as the permutation operator that exchanges the indices $i$ and $j$ as below,
\begin{equation}
    \label{eq:PermutationOperator}
    \begin{aligned}
    \Pi_{ij} \boldsymbol{e}_k =
    \begin{cases}
       \boldsymbol{e}_j \quad \text{for} \quad k = i, \\
       \boldsymbol{e}_i \quad \text{for} \quad k = j, \\
       \boldsymbol{e}_k \quad \text{for} \quad k \neq i,j.
    \end{cases}
    \end{aligned}
\end{equation}
the averaging matrix $\boldsymbol{W}_{i,j}$ can be written as
\begin{equation}
    \begin{aligned}
        \label{eq:AveragingMatrixInTermsOFPermutationOperator}
        \boldsymbol{W}_{i,j} = \left( \boldsymbol{I} + \Pi_{ij} \right)/2.
    \end{aligned}
\end{equation}
$\Pi_{ij}$ can be represented in the matrix form as below,
\begin{equation}
    \label{eq:PermutationOperatorMatrixForm}
    \begin{aligned}
       \Pi_{ij}  =  \boldsymbol{I}-(\boldsymbol{e}_i - \boldsymbol{e}_j) (\boldsymbol{e}_i - \boldsymbol{e}_j)^T.
    \end{aligned}
\end{equation}
The probabilities $P_{i,j}$ form a square matrix $\boldsymbol{P}$ that has the same sparsity pattern as the adjacency matrix of the graph $\mathcal{G}$.
In other words, $\boldsymbol{P}_{i,j}=0$ if ${i,j}\not\in E$.
Also all of its elements are nonnegative and each row of the matrix $\boldsymbol{P}$ should sum up to one, i.e. $\sum\nolimits_{j}^{}{\boldsymbol{P}_{i,j}}=1$.
This is due to the fact that when the clock in agent $i$ ticks, it has to communicate with one of its neighbors.
Based on the update equation (\ref{eq:Gossip1}) the state of agents at $k$-th time interval $(\boldsymbol{x}(k))$ can be related to the initial state of agents $(\boldsymbol{x}(0))$ as %
$\boldsymbol{x}(k) = \pmb{\phi}(k)\times \boldsymbol{x}(0)$,
where
$\pmb{\phi}(k) = \boldsymbol{W}(k)\times \boldsymbol{W}(k-1) \ldots \boldsymbol{W}(0)$.
Each one of matrices $\boldsymbol{W}(i)$ is an averaging matrix similar to $\boldsymbol{W}_{i,j}$ in (\ref{eq:Gossip2}).
The product $\pmb{\phi}(k)$ should converge to $\boldsymbol{1}\times \boldsymbol{1}^{T}/N$, in order to guarantee the convergence of $\boldsymbol{x}(k)$ to the average value or consensus state, in other words,
\begin{equation}
    \begin{aligned}
        \label{eq:Gossip3}
        \lim_{k \to \infty} \pmb{\phi}(k) = \left(\boldsymbol{1}\times \boldsymbol{1}^{T}\right)/N.
    \end{aligned}
\end{equation}
$\boldsymbol{1}$ is a column vector with $N$ elements, all equal to one.

Since the iteration (\ref{eq:Gossip1}) computes the average of states of agents then the sum of values of agents should be preserved, i.e.
$\boldsymbol{1}^T \times \boldsymbol{x}(k+1) = \boldsymbol{1}^{T} \times \boldsymbol{x}(k)$.
By substituting (\ref{eq:Gossip1}) we have
$\boldsymbol{1}^T \times \boldsymbol{W}(k) \times \boldsymbol{x}(k) = \boldsymbol{1}^{T} \times \boldsymbol{x}(k)$
This means that the vector of all ones $(\boldsymbol{1})$ is the left eigenvector of $\boldsymbol{W}(k)$ corresponding to eigenvalue one.
In addition, the vector of averages $\left( \frac{\boldsymbol{1}\times\boldsymbol{1}^{T}}{N} \right)\times x(0)$ must serve as the fixed point of iteration (\ref{eq:Gossip1}) i.e.
$\boldsymbol{W}(k)\times \frac{\boldsymbol{1}\times \boldsymbol{1}^{T}}{N} \times \boldsymbol{x}(0) = \frac{\boldsymbol{1}\times\boldsymbol{1}^{T}}{|{V}|} \times \boldsymbol{x}(0)$,
or in other words, the vector of all ones $(\boldsymbol{1})$ is the right eigenvector of $\boldsymbol{W}(k)$ corresponding to eigenvalue one.
Thus it can be deduced that $\boldsymbol{1}$ is the eigenvector corresponding to eigenvalue one for all averaging matrices that can serve as $\boldsymbol{W}(k)$ in iteration (\ref{eq:Gossip1}).

Let $P_i$ be
the probability that the clock in agent $i$ ticks at $k$-th time interval and $\boldsymbol{P}_{i,j}$ be the probability that agent $i$ communicates with agent $j$ given that its clock has ticked,
then the probability that $\boldsymbol{W}(k)$ is equal to the averaging matrix $\boldsymbol{W}_{i,j}$ is %
$P_i \cdot \boldsymbol{P}_{i,j}$ and the matrices $\boldsymbol{W}(k)$ are drawn independent and identically distributed (i.i.d.) from the set of possible averaging matrices.
Defining the gossip operator $(\overline{\boldsymbol{W}})$ as the mean of $\boldsymbol{W}(k)$ we have
\begin{equation}
    \begin{aligned}
        \label{eq:Gossip4}
        E\left( \pmb{\phi}(k) \right)  =  \prod\nolimits_{i=0}^{k}{E\left( \boldsymbol{W}(i) \right)}  =  \overline{\boldsymbol{W}}^{k}.
    \end{aligned}
\end{equation}

\subsection*{Non-Uniform Clock Distribution}
In the literature \cite{Xiao04Boyd} the probability that the clock in vertex $i$ ticks at $k$-th time interval ($P_i$) is assumed to be $1/N$ (i.e. uniform clock distribution) by default.
But here we have relaxed this constraint to enable a more general feasible region for the optimization problem and as a result a faster convergence rate for the gossip algorithm.
Uniform clock distribution is obtained by assuming that the clock at each vertex ticks according to a Poisson process where the Poisson processes in vertices have the same rate.
Thus, the inter-tick times at each vertex are rate $1$ exponentials, independent across vertices and over time.
This is equivalent to a single clock ticking according to a rate $N$ Poisson process.

The main reason that makes the Poisson process most suitable for the clock tick model of the gossip algorithm is the memoryless property of the Poisson process.
\subsubsection*{Memoryless Property of Poisson Process \cite{PoissonBook}}
Consider a clock that ticks according to a Poisson process with rate $\lambda$, we define the random variable $\gamma_{t}$ as the waiting time from time $t$ until the next clock tick.
Then for any $t \geq 0$, the random variable $\gamma_{t}$ has the same exponential distribution with mean $1/\lambda$, i.e. $P\{ \gamma_{t} \leq x \}  =  1  -  e^{-\lambda x}$, $x \geq 0$, independent of time $t$.

The memoryless property of Poisson process states that at each point in time, the waiting time until the next clock tick has the same exponential distribution as the original inter-tick time, regardless of how long ago the last clock tick occurred.
The Poisson process is the only renewal process having this memoryless property.
How much time is elapsed since the last clock tick gives no information about how long to wait until the next clock tick.
Another conclusion from this remarkable property of the Poisson process is that the number of clock ticks in any time interval of length $s$ has a Poisson distribution with mean $\lambda s$, independent of the beginning point of the time interval.
The following properties can be intuitively concluded from the memoryless property of the Poisson process.
\begin{itemize}
\item Independent increments: the numbers of clock ticks occurring in disjoint intervals of time are independent.
\item Stationary increments: the number of clock ticks occurring in a given time interval depends only on the length of the interval.
\item The probability of one clock ticking occurring in a time interval of length $\Delta t$ is $\lambda \Delta t + o(\Delta t)$ for $\Delta t \rightarrow 0$. %
\item The probability of two or more clock ticks occurring in a time interval of length $\Delta t$ is $o(\Delta t)$ for $\Delta t \rightarrow 0$.
\end{itemize}
The last property states that the probability of two or more clock ticks in a very small time interval of length $\Delta t$ is negligibly small compared to $\Delta t$ itself as $\Delta t \rightarrow 0$.

\subsubsection*{Merging of Poisson Processes}

The clock in each vertex ticks according to an independent Poisson process with rate $\lambda_i$, where
$N_{i}(t)$ is the number of clock ticks from $i$-th vertex up to time $t$.
In {\cite[Theorem 1.1.3]{PoissonBook}} it is shown that the merged process $\{N(t) = \sum_{i} {N_{i}(t)} \}$ for $t \geq 0$ is a poisson process with rate $\lambda = \sum_{j} {\lambda_j}$.
Denoting by $Z_{k}$ the time interval between the $(k - 1)$-th and $k$-th clock tick in the merged Poisson process and letting $I_k = i$ if the $k$-th clock tick in the merged Poisson process is the clock tick from $i$-th vertex, then the probability that the clock in vertex $i$ ticks at $k$-th time interval $(Z_{k})$ is
\begin{equation}
    \label{eq:ClockDistribution341}
    \begin{gathered}
        P_{i} = P\{  I_k = i | Z_{k} = t  \}  =   \frac{\lambda_i} {  \sum\nolimits_{j} { \lambda_j } },
    \end{gathered}
\end{equation}
which is independent of $t$.
Another perspective on (\ref{eq:ClockDistribution341}) is by considering the merged Poisson process $\{N(t), t \geq 0\}$.
Suppose that each clock tick in the merged Poisson process is the clock tick from $i$-th vertex with respective probabilities $P_{i}$, independently of the clocks in all other vertices.
Let $N_{i}(t)$ be the number of clock tick from $i$-th vertex up to time $t$.
Then $\{N_{i}(t), t \geq 0\}$ for $i=1, \ldots, N$ are $N$ independent Poisson processes having respective rates $λ P_{i}$.
The remarkable result (\ref{eq:ClockDistribution341}) states that the next clock tick is from the $i$-th vertex with probability
$\lambda_i / \sum_{j}{\lambda_{j}}$ regardless of how long it takes until the next arrival.
Based on the merging and the memoryless property of the Poisson process, it is apparent that the clock distribution of the gossip algorithm can be modified by setting the rate of the Poisson process in each vertex according to (\ref{eq:ClockDistribution341}).
In the final optimization stage we have provided the optimal answers for both cases of uniform and non-uniform clock and it is shown that the uniform case is suboptimal compared to the non-uniform one.
\subsection*{Convergence Rate \& Problem Formulation}
From (\ref{eq:Gossip4}) it can be deduced that in expectation convergence of $\pmb{\phi}(k)$ to $\boldsymbol{1}\times \boldsymbol{1}^{T}/N$ is equivalent to convergence of $\overline{\boldsymbol{W}}^{k}$ to $\boldsymbol{1}\times \boldsymbol{1}^{T}/N$.
In \cite{Xiao04Boyd} the following conditions have been stated as the necessary conditions for convergence of $\overline{\boldsymbol{W}}^{k}$ to $\boldsymbol{1}\times \boldsymbol{1}^{T}/N$ under the assumption of uniform selection of vertices.
\begin{equation}
    \begin{aligned}
        \label{eq:Gossip5}
        \boldsymbol{1}^{T} \times \overline{\boldsymbol{W}} = \boldsymbol{1}^{T},  \quad  \overline{\boldsymbol{W}} \times \boldsymbol{1} = \boldsymbol{1},
    \end{aligned}
\end{equation}
\begin{equation}
    \begin{aligned}
        \label{eq:Gossip7}
        \rho\left(\overline{\boldsymbol{W}} - \left(\boldsymbol{1}\times\boldsymbol{1}^{T}\right)/N \right)  <  1.
    \end{aligned}
\end{equation}
Here $\rho(.)$ refers to the spectral radius of a matrix.
It is straightforward to show that the same results holds true for the case of non-uniform selection of vertices.
In other words, assuming non-uniform selection of vertices, the conditions (\ref{eq:Gossip5}) and (\ref{eq:Gossip7}) are the necessary conditions for convergence of $\overline{\boldsymbol{W}}^{k}$ to $\boldsymbol{1}\times \boldsymbol{1}^{T}/N$.
The matrix $\overline{\boldsymbol{W}}$ satisfies conditions (\ref{eq:Gossip5}). % and (\ref{eq:Gossip6}).
This is independent of the distribution %$\left(\frac{1}{N}P_{i,j} \right)$
$\left(P_{i} \boldsymbol{P}_{i,j} \right)$ that each one of averaging matrices is selected as $\boldsymbol{W}(k)$.
This is due to the fact that $\overline{\boldsymbol{W}}$ is the expected value of averaging matrices (\ref{eq:Gossip2}) which all of them satisfy both conditions (\ref{eq:Gossip5}). % and (\ref{eq:Gossip6}).
Satisfying condition (\ref{eq:Gossip7}) relies solely on the probabilities $P_i$ and  $\boldsymbol{P}_{i,j}$.

Considering the averaging matrix (\ref{eq:Gossip2}), the gossip operator $(\overline{\boldsymbol{W}})$ can be written as
\begin{equation}
    \label{eq:GossipOperator}
    \begin{aligned}
       \overline{\boldsymbol{W}}  =  \sum\nolimits_{i,j} P_i \boldsymbol{P}_{ij}\boldsymbol{W}_{ij}.
    \end{aligned}
\end{equation}
In terms of the permutation operator $(\Pi_{ij})$
the gossip operator $(\overline{\boldsymbol{W}})$ can be written as %below
$ \overline{\boldsymbol{W}}=\frac{1}{2}(\sum_{i,j} P_i \boldsymbol{P}_{ij})2\boldsymbol{I}-\frac{1}{2}(\sum_{i,j} P_i \boldsymbol{P}_{ij})(\boldsymbol{I}-\Pi_{ij})  =  I-\frac{1}{2}\sum_{i,j} P_i \boldsymbol{P}_{ij}(\boldsymbol{I}-\Pi_{ij}) $.
The last equality is obtained from the fact that $\sum_{i,j} P_i \boldsymbol{P}_{ij}=\sum_i P_i (\sum_jP_{ij}=1)=1$.
Introducing $q_{ij}$ as
%\vspace{-2pt}
\begin{equation}
    \nonumber
    \begin{aligned}
       q_{ij}=\left(P_i \boldsymbol{P}_{ij} + P_j \boldsymbol{P}_{ji}\right)/2
    \end{aligned}
\end{equation}
The Gossip operator $\overline{\boldsymbol{W}}$ can be written in terms Of Laplacian operator $\boldsymbol{L}(q)$ as % below,
$\overline{\boldsymbol{W}} = \boldsymbol{I} - \boldsymbol{L}(q)$
with the Laplacian operator $\boldsymbol{L}(q)$ defined as
$\boldsymbol{L}(q)  =  \sum_{i,j} q_{ij}(\boldsymbol{I}-\Pi_{ij})$

To analyze the convergence rate of the gossip algorithm as described above, first we have to define the averaging time for the gossip algorithm.
\begin{definition} {Averaging Time}
\label{AveragingTimeGossip}
\\
For any $0<\epsilon<1$, the $\epsilon$-averaging time of the randomized gossip algorithm (denoted by $T_{ave}(\epsilon,\boldsymbol{P})$ is defined as follows:
\begin{equation}
    \begin{aligned}
        \nonumber
        T_{ave}(\epsilon,\boldsymbol{P})  =  \sup\limits_{\boldsymbol{x}(0)} \inf \left\{ k: \Pr{ \left( \frac { \| \boldsymbol{x}(k)-\boldsymbol{x}_{ave}\cdot\boldsymbol{1} \| } { \| \boldsymbol{x}(0) \| } \geq \epsilon \right) } \leq\epsilon \right\}.
    \end{aligned}
\end{equation}
$\boldsymbol{1}$ is the vector of all ones and $\|\boldsymbol{z}\|$ refers to the $l_{2}$ norm of the vector $\boldsymbol{z}$.
\end{definition}
In {\cite[Theorem~3]{Boyd06Gossip}} it is shown that the averaging time $(T_{ave}(\epsilon,\boldsymbol{P}))$ monotonically increases by second largest eigenvalue of % $\overline{\boldsymbol{W}}(P)=\sum\nolimits_{i,j=1}^{N}{\boldsymbol{P}_{i,j}\cdot \boldsymbol{W}_{i,j}/N}$.
$\overline{\boldsymbol{W}}(P)=\sum\nolimits_{i,j=1}^{N}{P_i \boldsymbol{P}_{i,j}\cdot \boldsymbol{W}_{i,j}}$.
Here the matrices $\boldsymbol{W}_{i,j}$ are the averaging matrices defined in (\ref{eq:Gossip2}).
Thus optimizing the convergence rate of the gossip algorithm is equivalent to minimizing the second largest eigenvalue of the matrix $\overline{\boldsymbol{W}}(\boldsymbol{P})$.
This minimization problem can be formulated as follows:
\begin{equation}
    \nonumber
    \begin{aligned}
        \min\limits_{\boldsymbol{P}_{i,j}} \quad &\lambda_{2}(\overline{\boldsymbol{W}}) \\
        s.t. \quad &\overline{\boldsymbol{W}}=\sum\nolimits_{i,j=1}^{N} {P_i \boldsymbol{P}_{i,j} \cdot \boldsymbol{W}_{i,j}}, \\
        &P_i \geq 0, \quad \boldsymbol{P}_{i,j} \geq 0,  \quad  \boldsymbol{P}_{i,j} = 0 \; \text{if} \; \{i,j\} \not\in \mathcal{E}, \\
        &\sum\nolimits_{i}^{} {\boldsymbol{P}_{i}} = 1, \quad \forall i: \sum\nolimits_{j}^{} {\boldsymbol{P}_{i,j}} = 1.
    \end{aligned}
\end{equation}
This is a convex optimization problem since the second largest eigenvalue of a doubly stochastic matrix is a convex function on the set of symmetric matrices.
This optimization problem can be reformulated as the following semidefinite programming problem \cite{Boyd06Gossip}.
\begin{equation}
    \label{eq:ClassicalGossipSDP}
    \begin{aligned}
        \min\limits_{P_i,\boldsymbol{P}_{i,j},s} \quad &s \\
        s.t. \quad &\overline{\boldsymbol{W}} - \frac{\boldsymbol{1}\times1^{T}}{N} \preceq s\boldsymbol{I},\\
        &\overline{\boldsymbol{W}} = \sum\nolimits_{i,j=1}^{N} {{P_i \boldsymbol{P}_{i,j} \cdot \boldsymbol{W}_{i,j}}}, \\
        &P_i \geq 0, \quad \boldsymbol{P}_{i,j} \geq 0,  \quad  \boldsymbol{P}_{i,j} = 0 \; \text{if} \; \{i,j\} \not\in \mathcal{E}, \\
        &\sum\nolimits_{i}^{} {\boldsymbol{P}_{i}} = 1, \quad \forall i: \sum\nolimits_{j}^{} {\boldsymbol{P}_{i,j}} = 1.
    \end{aligned}
\end{equation}
$\boldsymbol{B}\preceq \boldsymbol{A}$ means that the matrix $\boldsymbol{A} - \boldsymbol{B}$ is a positive semidefinite matrix.
We refer to problem (\ref{eq:ClassicalGossipSDP}) as the Fastest Classical Gossip Algorithm (FCGA) problem.

An automorphism of the graph $\mathcal{G} = (\mathcal{V}, \mathcal{E})$ is a permutation $\sigma$ of $\mathcal{V}$ such that $\{i,j\} \in \mathcal{E}$ if and only if $\{\sigma(i),\sigma(j)\}\in \mathcal{E}$, the set of all such permutations, with composition as the group operation, is called the automorphism group of the graph and denoted by $Aut(\mathcal{G})$.
For a vertex $i \in \mathcal{V}$, the set of all images $\sigma(i)$, as $\sigma$ varies through a subgroup $G \subseteq Aut(\mathcal{G})$, is called the orbit of $i$ under the action of $G$.
The vertex set $\mathcal{V}$ can be written as disjoint union of distinct orbits.
In \cite{Ghosh06Boyd}, it has been shown that the optimal probabilities on the edges within an orbit are equal.

\section{Quantum Gossip Algorithm}
\label{sec:QuantumGossipAlgorithm}

We consider a quantum network as a composite (or multipartite) quantum system with $N$ qudits.
Assuming $\mathcal{H}$ as the d-dimensional Hilbert space over $\mathbb{C}$, then the state space of the quantum network is within the Hilbert space $\mathcal{H}^{\otimes N} = \mathcal{H} \otimes \ldots \otimes \mathcal{H}$.
The state of the quantum system is described by its density matrix $(\boldsymbol{\rho})$.
This matrix is positive Hermitian and its trace is one $(tr(\boldsymbol{\rho}) = 1)$.
The network is associated with an underlying graph $\mathcal{G}=\{ \mathcal{V}, \mathcal{E} \}$, where $\mathcal{V}=\{1,\ldots, N\}$ is the set of indices for the $N$ qudits, and each element in $\mathcal{E}$ is an unordered pair of two distinct qudits, denoted as $\{j,k\} \in \mathcal{E}$ with $j,k \in \mathcal{V}$.
Permutation group $S_{N}$ acts in a natural way on $\mathcal{V}$ by mapping $\mathcal{V}$ onto itself.
For each permutation $\pi \in S_{N}$ %in $\mathcal{V}$
we associate unitary operator $U_{\pi}$ over $\mathcal{H}^{\otimes N}$, as below
\begin{equation}
    \nonumber
    \begin{gathered}
        U_{\pi} ( Q_{1} \otimes \cdots \otimes Q_{N} ) = Q_{\pi(1)} \otimes \cdots \otimes Q_{\pi(N)},
     \end{gathered}
\end{equation}
where $Q_{i}$ is an operator in $\mathcal{H}$ for all $i = 1, \ldots, N$.
A special case of permutations is the swapping permutation or transposition where $\pi(j)=k$, $\pi(k)=j$ and $\pi(i) = i$ for all $i \in \mathcal{V}$ and $i \notin {j,k}$
We denote the swapping permutation between the qudits indices $j$ and $k$ by $\pi_{j,k}$ and the corresponding swapping operator by $U_{j,k}$.
% The swapping operator $U_{j,k}$ in terms of the generalized Gell-Mann matrices is provided in Appendix \ref{sec:GellMannMatrices}.
In {\cite[Appendix~A]{SaberQConsensusContinuous}} the swapping operator $U_{j,k}$ has been expressed as linear combination of the Cartesian product of Gell-Mann matrices.

In the model of quantum network presented here, every vertex represents a particle.
Employing the quantum gossip interaction introduced in \cite{PetersenRef15,JohanssonFiniteTimeConvergentGossiping},
on each iteration of the gossip algorithm,
one particle (say $j$-th particle) is selected with probability % $1/N$.
$P_j$.
Then this particle selects one of its neighbours for updating its state.
This neighbouring particle is selected with probability $\boldsymbol{P}_{j,k}$ according to the underlying graph of the quantum network,
and the updating procedure on the states of $j$-th and $k$-th particles can be written
in terms of the evolution of the density matrix as below,
\begin{equation}
    \label{eq:Lindblad2}
    \begin{gathered}
        \boldsymbol{\rho(t+1)}=\frac{1}{2}(\boldsymbol{\rho(t)}+U_{j,k}\boldsymbol{\rho(t)}U^\dagger_{j,k})=\boldsymbol{\rho(t)}+\frac{1}{2}(-\boldsymbol{\rho(t)}+U_{j,k}\boldsymbol{\rho(t)}U^\dagger_{j,k}),
    \end{gathered}
\end{equation}
where its resultant quantum consensus state % \cite{PetersenRef15}
is as below
\begin{equation}
    \label{eq:QCMEFinalConsensus}
    \begin{gathered}
        \boldsymbol{\rho}^{*}  = \frac{1}{N!}  \sum_{\pi \in S_{N}} {U_{\pi} \boldsymbol{\rho}(0) U_{\pi}^{\dagger} }   .
     \end{gathered}
\end{equation}
A necessary condition for the quantum gossip algorithm to reach its quantum consensus state (i.e. $\lim_{t \rightarrow \infty} {\boldsymbol{\rho}}(t)  =  \boldsymbol{\rho}^{*}$) is that the underlying graph of the quantum network should be connected.

The analysis presented in this paper aim at optimizing the convergence rate of the quantum gossip algorithm to its quantum consensus state.
To this aim, first we expand the density matrix $(\boldsymbol{\rho})$ as the linear combination of the generalized Gell-Mann matrices {\cite[Appendix~A]{SaberQConsensusContinuous}} as below,
\begin{equation}
    \label{eq:DecompositionDensityGeneral}
    \begin{gathered}
        \boldsymbol{\rho} = \frac{1}{d^{2N}} \sum_{ \mu_{1}, \mu_{2}, \ldots, \mu_{N} = 0 }^{ d^{2} - 1 }  {  \rho_{ \mu_{1}, \mu_{2}, \ldots, \mu_{N} } \cdot \lambda_{\mu_{1}} \otimes  \lambda_{\mu_{2}} \otimes \cdots \otimes \lambda_{\mu_{N}}  },
     \end{gathered}
\end{equation}
where $N$ is the number of particles and $\otimes$ denotes the Cartesian product and $\lambda$ matrices are the generalized Gell-Mann matrices {\cite[Appendix~A]{SaberQConsensusContinuous}}.
Due to Hermity of density matrix, its coefficients of expansion $\rho_{ \mu_{1}, \mu_{2}, \ldots, \mu_{N} }$ are real numbers and because of unit trace of $\boldsymbol{\rho}$ we have $\rho_{0,0,\ldots,0} = 1$.
Based on the decomposition of $\boldsymbol{\rho}$ (\ref{eq:DecompositionDensityGeneral}), its permutations can be written as % below
\begin{equation}
    \label{eq:DecompositionDensityPermutation}
    \begin{gathered}
        U_{j,k} \times \boldsymbol{\rho} \times U_{j,k}^{\dagger} =
        \frac{1}{2^N} \sum_{ \mu_{1}, \ldots \mu_{N} = 0 }^{ d^{2} - 1 }  {  \rho_{ \mu_{1}, \ldots \mu_{k}, \ldots, \mu_{j}, \ldots, \mu_{N} } \cdot \lambda_{\mu_{1}} \otimes \cdots \lambda_{\mu_{j}} \otimes \cdots \lambda_{\mu_{k}} \otimes \cdots \otimes \lambda_{\mu_{N}}  }
     \end{gathered}
\end{equation}
Note that in (\ref{eq:DecompositionDensityPermutation}) due to permutation operators, the place of indices $\mu_{j}$ and $\mu_{k}$ in the index of parameter $\boldsymbol{\rho}$ are interchanged.
Substituting the density matrix $\boldsymbol{\rho}$ from (\ref{eq:DecompositionDensityGeneral}) and its permutation (\ref{eq:DecompositionDensityPermutation}) in (\ref{eq:Lindblad2}) and considering the independence of the matrices $\lambda_{\mu_{1}} \otimes \lambda_{\mu_{2}} \otimes \cdots \lambda_{\mu_{N}} $, we can conclude the following for the evolution of the density matrix (\ref{eq:Lindblad2}),
\begin{equation}
    \label{eq:DensityEquation1}
    \begin{gathered}
        \rho_{\mu_{1}, \ldots, %\mu_{j}, \cdots, \mu_{k}, \cdots,
        \mu_{N}}(t+1)  =   \frac{1}{2}( \rho_{\mu_{1},\ldots,\mu_{k},\ldots,\mu_{j},\ldots,\mu_{N} }(t)  +  \rho_{\mu_{1},\ldots,\mu_{j},\ldots,\mu_{k},\ldots,\mu_{N} }(t) )
    \end{gathered}
\end{equation}
for all $\mu_{1},\mu_{2},\cdots,\mu_{N}=0,\cdots,d^{2}-1,$.
Following the same procedure, the tensor component of the consensus state of the quantum gossip algorithm (\ref{eq:QCMEFinalConsensus}) can be written as below
\begin{equation}
    \label{eq:QuantumConsensusState872}
    \begin{gathered}
        \rho_{\mu_1, \mu_2, \ldots, \mu_N}^{*}  =  \frac{1}{N!}  \sum_{\pi \in S_{N}} {\rho_{\pi(\mu_1), \pi(\mu_2), \ldots, \pi(\mu_N)} (0)}
    \end{gathered}
\end{equation}
and for the connected underlying graph, the quantum gossip algorithm reaches quantum consensus, componentwise as $\lim_{t \rightarrow \infty} {    \rho_{\mu_1, \mu_2, \ldots, \mu_N}  (t)    }  =  \rho_{\mu_1, \mu_2, \ldots, \mu_N}^{*}$.
Defining $\boldsymbol{X}_Q$ as a column vector of length $d^{2N}$ with components $\rho_{\mu_1, \ldots, \mu_N}$, the evolution of the density matrix (\ref{eq:Lindblad2}) can be written as blow,
\begin{equation}
    \label{eq:QuantumStateUpdate}
    \begin{gathered}
        \boldsymbol{X}_Q(t+1)  =  \frac{\boldsymbol{I}_{d^{2N}}+ U_{j,k} } {2} \cdot \boldsymbol{X}_Q(t)  .
    \end{gathered}
\end{equation}
where $U_{j,k}$ is the swapping operator {\cite[Appendix~A]{SaberQConsensusContinuous}}, provided that $d$ is replaced with $d^2$ which in turn results in Gell-Mann matrices of size  $d^2 \times d^2$.
Comparing the set of equations in (\ref{eq:DensityEquation1}) with those of the classical gossip in (\ref{eq:Gossip1}) we can see that the quantum gossip (\ref{eq:QuantumStateUpdate}) is transformed into the classical gossip (\ref{eq:Gossip1}) with $d^{2N} -1$ tensor component $\rho_{\mu_{1}, \cdots, \mu_{N}}$ as the agents' states.
Thus, similar to the analysis performed in section \ref{sec:ClassicalGossip} for the classical gossip algorithm, it is straight forward to see that the convergence rate of the quantum gossip algorithm is governed by the second largest eigenvalue of the quantum gossip operator $\overline{\boldsymbol{W}}_{Q}(\boldsymbol{P})$ defined as
\begin{equation}
    \label{eq:MeanAveragingMatricesQuantum}
    \begin{gathered}
        \overline{\boldsymbol{W}}_{Q}(\boldsymbol{P})  =  \sum_{\{j,k\}\in \mathcal{E}}\left(P_j  \boldsymbol{P}_{jk} + P_k  \boldsymbol{P}_{kj}\right) \cdot \frac{ \boldsymbol{I}_{d^{2N}} + U_{j,k} } {2}.
    \end{gathered}
\end{equation}
The quantum gossip operator $\overline{\boldsymbol{W}}_{Q}(\boldsymbol{P})$ is obtained by taking the average over all possible swapping (according to the underlying graph) where
the probability that a particular pair $(j,k)$ is swapped at time $t$ is $P_j \cdot \boldsymbol{P}_{j,k}  +   P_k \cdot \boldsymbol{P}_{k,j}$.
The quantum gossip operator $(\overline{\boldsymbol{W}}_{Q}(\boldsymbol{P}))$ can be written in terms of quantum Laplacian operator $\boldsymbol{L}_{Q}(q)$ as %below,
$\overline{\boldsymbol{W}}_{Q}(\boldsymbol{P})  =   \boldsymbol{I}  -  \boldsymbol{L}_{Q}(q)$, where $\boldsymbol{L}_{Q}(q)$ can be written as
$\boldsymbol{L}_Q (q)  =  \sum_{\{j,k\} \in \mathcal{E}} {  \boldsymbol{q}_{j,k} ( \boldsymbol{I}_{d^{2N}}  -  U_{j,k} )  }=\sum_{\{j,k\} \in \mathcal{E}}\left(P_j \cdot \boldsymbol{P}_{j,k} + P_k \cdot \boldsymbol{P}_{k,j}\right)( \boldsymbol{I}_{d^{2N}}  -  U_{j,k} )$.

Based on (\ref{eq:MeanAveragingMatricesQuantum}) and the analysis presented in section \ref{sec:ClassicalGossip} for the classical gossip algorithm, it can be concluded that optimizing the convergence rate of the quantum gossip algorithm is equivalent to minimizing the second largest eigenvalue of the quantum gossip operator $(\overline{\boldsymbol{W}}_{Q}(\boldsymbol{P}))$.
This minimization problem can be formulated as following convex optimization problem,
\begin{equation}
    \label{eq:QuantumGossipConvexFormulation}
    \begin{aligned}
        \min\limits_{\boldsymbol{P}_{i,j}} \; &\lambda_{2}(\overline{\boldsymbol{W}}_{Q}(\boldsymbol{P})) \\
        s.t. \; &\overline{\boldsymbol{W}}_{Q}(\boldsymbol{P})   =    \sum\nolimits_{i,j=1}^{N}{\left(P_i \cdot \boldsymbol{P}_{ij}  + P_j \cdot \boldsymbol{P}_{ji}\right)\cdot \frac{\boldsymbol{I}_{d^{2N}}+ U_{i,j} } {2}} \\
        &P_i \geq 0, \quad \boldsymbol{P}_{i,j} \geq 0,  \quad  \boldsymbol{P}_{i,j} = 0 \; \text{if} \; \{i,j\} \not\in \mathcal{E}, \\
        &\sum\nolimits_{i}^{} {P_{i}} = 1,   \quad   \forall i: \sum\nolimits_{j}^{} {\boldsymbol{P}_{i,j}} = 1.
    \end{aligned}
\end{equation}
(\ref{eq:QuantumGossipConvexFormulation}) is referred as the Fastest Quantum Gossip (FQG) problem.

The underlying graph of classical gossip algorithm corresponding to (\ref{eq:QuantumStateUpdate}) is a cluster of connected components where each connected graph component corresponds to a given partition of $N$ into $K$ integers, namely $N = n_{1} + n_{2} + \cdots + n_{K}$, where $K \leq d^2$ and $n_{j}$ for $j=1,\ldots,K$  is the number of indices in $\rho_{\mu_1, \mu_2, \ldots, \mu_N}$ with equal values.
The connected graph components are referred to as the induced graphs.
The induced graphs originate from the tensor components $\rho_{\mu_{1}, \cdots, \mu_{N}}$ that can be transformed into each other by permuting their indices and thus  updating their states to their average value according to (\ref{eq:DensityEquation1}).
By establishing the intertwining relation between the induced graphs in \cite{SaberQConsensusContinuous}, the results of Aldous's conjecture is generalized to the set of all induced graphs and it is shown that the second eigenvalue of the Laplacian matrices of all the induced graphs are the same.
Therefore for optimizing the convergence rate of the corresponding classical gossip algorithm (\ref{eq:QuantumGossipConvexFormulation}), it is enough to optimize the problem for the induced graph corresponding to partition $(N-1,1)$ which is same as the underlying graph of the quantum network.
In \cite{SaberQConsensusDiscrete} similar results regarding generalization of Aldous's conjecture has been achieved by exploiting the Specht module representation of partitions of $N$.
%

%%%%%%%%%%%%%%%%%%%%%%%%%%%%%%%%%%%%%%%%%
%
%
%
%
%
%

%
\section{Optimization of the Convergence Rate}
\label{sec:OptimizationSectionComplete}

In this section we address the optimization of the Fastest Classical Gossip Algorithm (FCGA) problem (\ref{eq:ClassicalGossipSDP}) % Fastest Quantum Gossip (FQG) problem (\ref{eq:QuantumGossipConvexFormulation}),
over different network topologies.
In the first two subsection, we represent the optimal results for all possible topologies with $N = 4$ vertices which are connected and non-isomorphic and the topologies which the FCGA problem (\ref{eq:ClassicalGossipSDP}) can be solved using linear programming.
In the third and fourth subsections, the optimization of the FCGA problem (\ref{eq:ClassicalGossipSDP}) is addressed using Semidefinite Programming (SDP) for uniform and non-uniform clock distributions, respectively.
In the analysis presented in this section,
the complete solution procedure is provided only for the symmetric star topology.
For the rest of topologies we have provided the final results i.e. the optimal probabilities and the optimal value of the second largest eigenvalue of the gossip operator $(\overline{\boldsymbol{W}})$.
The notations and derivations of the semidefinite programming used in this section % and section \ref{sec:OptimizationSectionNonUniform}
are adopted from  {\cite[Appendix~B]{SaberQConsensusContinuous}}.

\subsection{Topologies with $N = 4$ vertices}

\begin{figure}
  \centering
     \includegraphics[width=160mm]{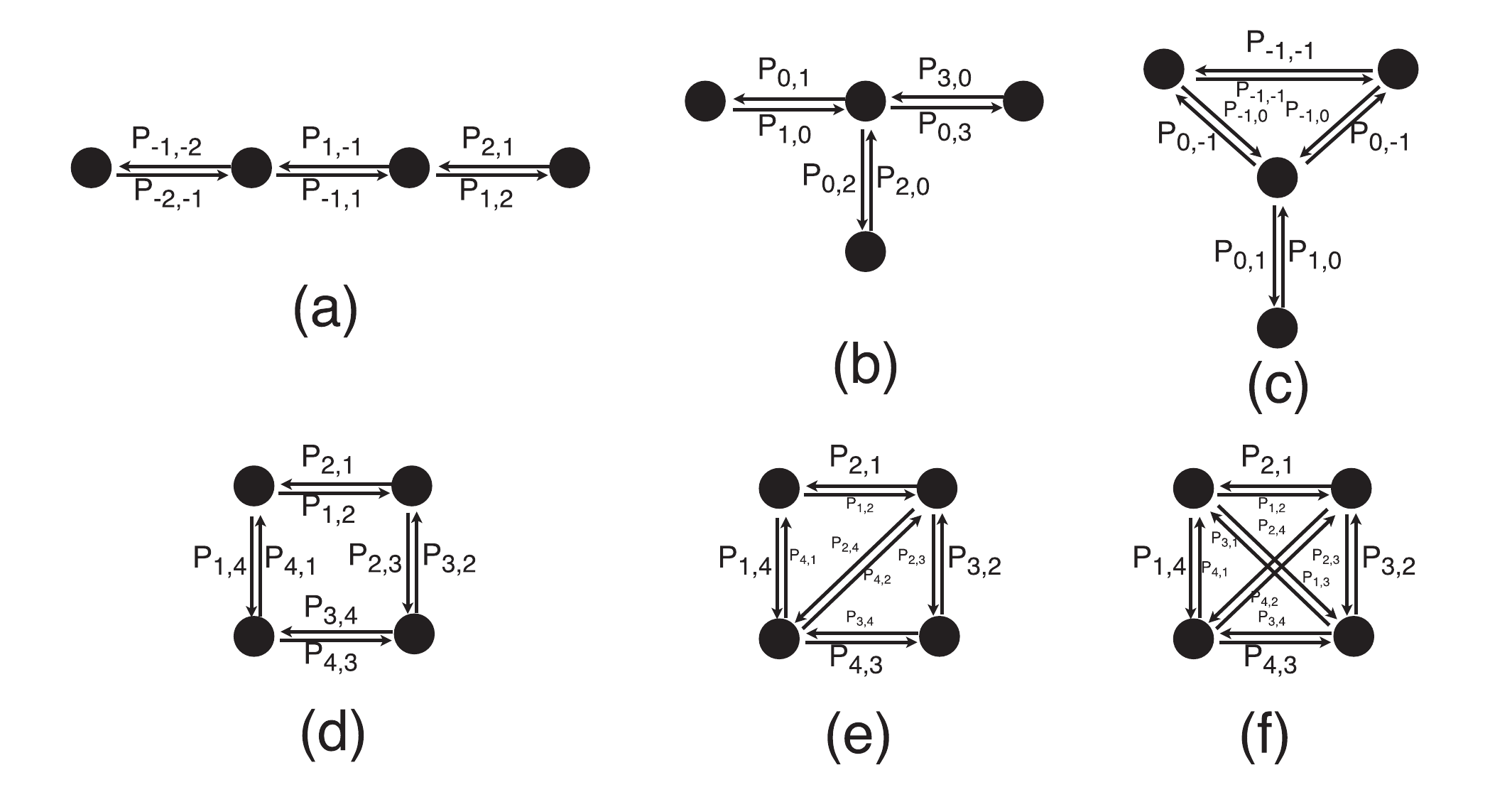}
  \caption{All possible connected underlying topologies with $N=4$ vertices which are non-isomorphic. (a) Path graph, (b) Star graph, (c) Lollipop graph, (d) Cycle graph, (e) Paw graph, (f) Complete graph.}
  \label{fig:N4Graphs}
\end{figure}

Here we provide the optimal results for all possible topologies with $N = 4$ vertices which are connected and non-isomorphic.
In Table \ref{tab:N4SLEMOptimal}, all possible connected topologies with $N=4$ vertices (as depicted in figure \ref{fig:N4Graphs}) are listed along with their optimal transition and clock distribution probabilities along with the optimal value of the second largest eigenvalue of the gossip operator $\lambda_2(\overline{\boldsymbol{W}})$.
Note that for all topologies listed in Table \ref{tab:N4SLEMOptimal}, the optimal value of the second largest eigenvalue of the gossip operator $\left(\lambda_2(\overline{\boldsymbol{W}})\right)$ obtained for uniform clock distribution is same as that of the non-uniform clock distribution.

\begin{table}[hb]
    \centering
    \caption{All non-isomorphic connected topologies with $N=4$ vertices with their optimal probabilities and the optimal value of the second largest eigenvalue of the gossip operator $\lambda_2(\overline{\boldsymbol{W}})$.}
    \label{tab:N4SLEMOptimal}
    \begin{tabular}{|c|c|c|c|} \hline
    {\hspace{-8pt}\begin{minipage}{0.30in}Topology\end{minipage}}
    &{  \hspace{-5pt}$\lambda_2(\overline{\boldsymbol{W}})$\hspace{-6pt}}
    &{  \hspace{-15pt}  \begin{minipage}{1.50in} \vspace{1pt} \hspace{1pt}Clock  Distribution \\ Probabilities \vspace{1pt} \end{minipage}   \hspace{-15pt}   }
    &{  \hspace{-17pt} \begin{minipage}{1.50in} \hspace{1pt} Transition  Probabilities\end{minipage} }  \\ \hline
    {\hspace{-6pt}  \begin{minipage}{0.30in}Path\end{minipage}  }
    &{  \hspace{-5pt} $9/10$ \hspace{-5pt} }
    &{  \hspace{-14pt}   \begin{minipage}{1.50in}  $\boldsymbol{P}_{-2} = \boldsymbol{P}_{-1} = 1/4$  \\    $\boldsymbol{P}_{1} = \boldsymbol{P}_{2} = 1/4$ \end{minipage}  \hspace{-15pt}  }
    &{   \hspace{-7pt}    \begin{minipage}{1.50in} \vspace{1pt} $\boldsymbol{P}_{-2,-1} = \boldsymbol{P}_{2,1} = 1$  \\  $\boldsymbol{P}_{-1,-2} = \boldsymbol{P}_{1,2} = 1/5$  \\  $\boldsymbol{P}_{1,-1} = \boldsymbol{P}_{-1,1} = 4/5$  \vspace{1pt} \end{minipage}   }  \\  \hline
    {   \hspace{-6pt}   \begin{minipage}{0.30in}Star\end{minipage}  }
    &{  \hspace{-5pt}  $5/6$  \hspace{-5pt}  }
    &{  \hspace{-14pt}   \begin{minipage}{1.50in}   $\boldsymbol{P}_{0} = \boldsymbol{P}_{1} = 1/4$  \\  $\boldsymbol{P}_{2} = \boldsymbol{P}_{3} = 1/4$     \end{minipage}  \hspace{-15pt}  }
    &{ \hspace{-7pt}  \begin{minipage}{1.50in}   \vspace{1pt}   $\boldsymbol{P}_{1,0} = \boldsymbol{P}_{2,0} = \boldsymbol{P}_{3,0} = 1$  \\  $\boldsymbol{P}_{0,1} = \boldsymbol{P}_{0,2} = \boldsymbol{P}_{0,3} = \frac{1}{3}$   \vspace{1pt}  \end{minipage}   }  \\  \hline
    {   \hspace{-6pt} \begin{minipage}{0.30in}Lollipop\end{minipage}   }
    &{  \hspace{-12pt} $\frac{3+\sqrt{3}}{4+\sqrt{3}}$  \hspace{-11pt} }
    &{  \hspace{-14pt}   \begin{minipage}{1.50in}  $\boldsymbol{P}_{-2} = \boldsymbol{P}_{-1} = 1/4$   \\    $\boldsymbol{P}_0 = 1/4$  \\  $\boldsymbol{P}_{1} = 1/4$    \end{minipage}  \hspace{-15pt}  }
    &{ \hspace{-7pt}   \begin{minipage}{1.50in}   \vspace{1pt}   $\boldsymbol{P}_{0,1} = \frac{5+2\sqrt{3}}{13}, \; \boldsymbol{P}_{1,0} = 1$    \\    $\boldsymbol{P}_{0,-1} = \frac{4 - \sqrt{3}}{13}$, \\  $\boldsymbol{P}_{-1,0} = \frac{24 + 7\sqrt{3}}{39}$  \\  $\boldsymbol{P}_{-1,-1} = \frac{15 - 7\sqrt{3}}{39}$  \vspace{1pt} \end{minipage}   }  \\ \hline
    {  \hspace{-6pt}  \begin{minipage}{0.30in}Cycle\end{minipage}   }
    &{  \hspace{-14pt}  $3/4$ \hspace{-11pt}    }
    &{  \hspace{-14pt}  \begin{minipage}{1.50in}  $\boldsymbol{P}_{1} = \boldsymbol{P}_{2} = 1/4$   \\ $\boldsymbol{P}_{3} = \boldsymbol{P}_{4} = 1/4$    \end{minipage}    \hspace{-15pt}  }
    &{ \hspace{-7pt}   \begin{minipage}{1.50in}  \vspace{1pt}  $\boldsymbol{P}_{i,i+1} = \boldsymbol{P}_{i+1,i}  = 1/2$   \\  for $i=1,2,3$  \\ $\boldsymbol{P}_{1,4} = \boldsymbol{P}_{4,1}  = 1/2$  \vspace{1pt}  \end{minipage}    }   \\ \hline
    {  \hspace{-4pt} \begin{minipage}{0.30in}  Paw  \end{minipage} }
    &{  \hspace{-14pt}  $3/4$  \hspace{-11pt}     }
    &{  \hspace{0pt}  \begin{minipage}{1.50in}  $\boldsymbol{P}_{1} = \boldsymbol{P}_{2} = 1/4$   \\   $\boldsymbol{P}_{3} = \boldsymbol{P}_{4} = 1/4$    \end{minipage}    \hspace{-15pt}  }
    &{  \hspace{-7pt}  \begin{minipage}{1.50in}  \vspace{1pt}  $\boldsymbol{P}_{i,i+1} = \boldsymbol{P}_{i+1,i}  = 1/2$   \\  for $i=1,2,3$  \\ $\boldsymbol{P}_{1,4} = \boldsymbol{P}_{4,1}  = 1/2$  \\  $\boldsymbol{P}_{4,2} = \boldsymbol{P}_{2,4} = 1/2$   \vspace{1pt}  \end{minipage}    }  \\ \hline
    {   \hspace{-5pt} \begin{minipage}{0.9in}Complete Graph\end{minipage}   }
    &{  \hspace{-14pt}$2/3$\hspace{-11pt}    }
    &{  \hspace{-14pt} \begin{minipage}{1.50in} \hspace{3pt} $\boldsymbol{P}_{i} = 1/4$   for $i=1,..,4$ \end{minipage}   \hspace{-15pt}  }
    &{  \hspace{-7pt}  \begin{minipage}{1.50in}  \vspace{1pt} $\boldsymbol{P}_{i,j}=1/3$   for $i,j=1,..,4$ \vspace{1pt} \end{minipage}    }  \\  \hline
    \end{tabular}
\end{table}

\subsection{Optimal Results Obtained through Linear Programming}
In this subsection, we provide the optimal results for a number of topologies where the FCGA problem (\ref{eq:ClassicalGossipSDP}) can be solved using Linear Programming \cite{BoydConvexBook}.
Note that the results reported in this subsection (other than that of Wheel topology) are globally optimal and they are obtained for uniform clock distribution.
In other words, optimizing the FCGA problem (\ref{eq:ClassicalGossipSDP}) with non-uniform clock distribution over topologies (other than Wheel topology) discussed in this subsection would result in the same convergence rate obtained from solving the FCGA problem (\ref{eq:ClassicalGossipSDP}) with uniform clock distribution.
In case of the Wheel topology non-uniform clock distribution would result in the global optimal answer.

\subsubsection{Cartesian Product of Edge Transitive Graphs}
This topology is obtained from Cartesian product of $m$ edge-transitive graphs.
The Laplacian operator $(\boldsymbol{L}(q))$ for the whole graph can be written as below,
\begin{equation}
    \nonumber
    \begin{gathered}
        \boldsymbol{L}(q) = \sum\nolimits_{i=1}^{m}     {   \boldsymbol{I}_{N_{1}} \otimes \boldsymbol{I}_{N_{2}} \otimes \cdots \otimes \boldsymbol{I}_{N_{i-1}} \otimes \boldsymbol{L}_{i}(q) \otimes \boldsymbol{I}_{N_{i+1}} \otimes \cdots \otimes \boldsymbol{I}_{N_{m}}       }
    \end{gathered}
\end{equation}
where $\boldsymbol{L}_{i}(q)$ is the Laplacian operator of $i$-th graph.
$\boldsymbol{I}_{N_{j}}$ is the identity matrix with size $N_{j}$ and $N_{j}$ is the number of vertices in the $j$-th graph.
Due to edge-transitivity, all edges of each edge-transitive graph have the same optimal transition probabilities.
Thus the Laplacian operator $(\boldsymbol{L}_{i}(q))$ for each one of the graphs can be written as $\boldsymbol{L}_{i}(q) = q_i \cdot \boldsymbol{L}_{i}$ in terms of its unweighted Laplacian matrix $(\boldsymbol{L}_{i})$ of the $i$-th graph.
Using this relation the Laplacian operator for the whole graph can be derived as below,
\begin{equation}
    \label{eq:CartesianLaplacian}
    \begin{gathered}
        \boldsymbol{L}(q) = \sum\nolimits_{i=1}^{m}     {   q_i \cdot \boldsymbol{I}_{N_{1}} \otimes  \cdots \boldsymbol{I}_{N_{i-1}} \otimes \boldsymbol{L}_{i}(q) \otimes \boldsymbol{I}_{N_{i+1}}  \cdots \otimes \boldsymbol{I}_{N_{m}}       }
    \end{gathered}
\end{equation}
We denote the eigenvalues of the $i$-th unweighted Laplacian matrix $\boldsymbol{L}_{i}$ in their sorted order by $\lambda_{i,\alpha_i}$ where $\alpha_i$ varies from $1$ to $N_i$.
Using this notation the eigenvalues of the Laplacian operator of the whole graph can be written as below,
\begin{equation}
    \label{eq:CartesianLaplacianEigenvalue}
    \begin{gathered}
        \lambda^{q}_{\alpha_1, \alpha_1, \ldots, \alpha_m}  =  q_1 \cdot \lambda_{1,\alpha_1}  +   q_2 \cdot \lambda_{2,\alpha_2}  +  \cdots  + q_m \cdot \lambda_{m,\alpha_m}
    \end{gathered}
\end{equation}
where $\alpha_i$ for $i=1,\ldots,m$ varies from $1$ to $N_i$.
Based on the derivation in (\ref{eq:CartesianLaplacianEigenvalue}) and considering the fact that the first eigenvalue of each unweighted Laplacian matrix $\boldsymbol{L}_{i}$ is zero (i.e. $\lambda_{i,1} = 0$ for $i = 1, \ldots, m$), the second smallest eigenvalue of the Laplacian operator of the whole graph can be written as $\lambda^{q}_2  =  \min { q_1 \cdot \lambda_{1,2},  q_2 \cdot \lambda_{2,2}, \cdots,   q_m \cdot \lambda_{m,2}  }$.
Using this result the optimization problem for the FCGA problem can be written as below,
\begin{equation}
    \label{eq:CartesianLP}
    \begin{aligned}
        \max\limits_{q_1, q_2, \cdots, q_m} \quad &s=\min_{i}{q_i\cdot\lambda_{i,2}},  \\
        s.t.  \quad\quad \   &\sum\nolimits_{j=1}^{m}{\tilde{E}_j \cdot q_j} = 1/2.
    \end{aligned}
\end{equation}
where $\tilde{E}_j = E_j \cdot \prod_{\substack{k=1\\k\neq j}}^{m} {N_{k}}$ and $E_j$ is the number of edges in the $j$-th edge-transitive graph.
For the optimal answer we have
\begin{equation}
    \label{eq:CartesianLPOptimalRelaion}
    \begin{aligned}
        s = q_1 \cdot \lambda_{1,2}  =  q_2 \cdot \lambda_{2,2}  =  \cdots  =  q_m \cdot \lambda_{m,2}.
    \end{aligned}
\end{equation}
From this relation we can conclude the following for the optimal value of the second largest eigenvalue of the gossip operator $\lambda_2(\overline{\boldsymbol{W}})$ and the transition probabilities
\begin{equation}
    \label{eq:CartesianLPOptimalSLEM}
    \begin{gathered}
        \lambda_2(\overline{\boldsymbol{W}})  =  1-  \frac{1}{2N \cdot \sum\nolimits_{j=1}^{m}{\frac{E_j} {N_j \cdot \lambda_{j,2} }} },
     \end{gathered}
\end{equation}
\begin{equation}
    \label{eq:CartesianLPOptimalprobability}
    \begin{aligned}
        &\boldsymbol{P}_{(\eta_1,\eta_2,\cdots,\eta_j,\cdots, \eta_m) \rightarrow (\eta_1,\eta_2,\cdots,\eta_j+1,\cdots, \eta_m) }  = \\
        &\qquad \qquad \qquad \qquad \boldsymbol{P}_{(\eta_1,\eta_2,\cdots,\eta_j+1,\cdots, \eta_m)\rightarrow (\eta_1,\eta_2,\cdots,\eta_j,\cdots, \eta_m) } =  \\
        &\qquad \qquad \qquad \qquad \qquad \qquad \qquad \qquad \frac{1} {2\lambda_{j,2}\left( \sum\nolimits_{j=1}^{m}{\frac{E_j}{N_j \cdot \lambda_{j,2} }} \right)}  ,\; j=1, \ldots, m,
     \end{aligned}
\end{equation}
where $N = \prod_{i=1}^{m}{N_{i}}$.
In (\ref{eq:CartesianLPOptimalprobability}), we have used the notation $(\eta_1,\cdots,\eta_j,\cdots, \eta_m)$ to refer to each vertex in the topology, where $\eta_1$ varies from $1$ to $N_i$.
using this notation, $\boldsymbol{P}_{(\eta_1,\cdots,\eta_j,\cdots, \eta_m) \rightarrow (\eta_1,\cdots,\eta_j+1,\cdots, \eta_m) }$ is the transition probability from vertex $(\eta_1,\cdots,\eta_j,\cdots, \eta_m)$  to  vertex  $(\eta_1,\cdots,\eta_j+1,\cdots, \eta_m)$.

As an example consider the Cartesian product of two complete graphs each with $N_1$ and $N_2$ vertices.
Let $q_1$ and $q_2$ be defined over the edges of each one of the complete graphs, then for the optimal results we have
\begin{subequations}
    \label{eq:CartesianCompleteGraphOptimalAnswer}
    \begin{gather}
        \lambda_2(\overline{\boldsymbol{W}})  =  1- \frac{1}     {   2 N_1 N_2 - N_1 - N_2   },      \label{eq:CartesianCompleteGraphOptimalAnswerS} \\
        \boldsymbol{P}_{(\eta_1,\eta_2)\rightarrow (\eta_1+1,\eta_2)}= \boldsymbol{P}_{(\eta_1+1,\eta_2)\rightarrow (\eta_1,\eta_2)}  =  \frac { N_2 }   { 2 N_1 N_2 - N_1 - N_2   },            \\
        \boldsymbol{P}_{(\eta_1,\eta_2)\rightarrow (\eta_1,\eta_2+1)}= \boldsymbol{P}_{(\eta_1,\eta_2+1)\rightarrow (\eta_1,\eta_2)}  = \frac {N_1}  {   2 N_1 N_2 - N_1 - N_2   },   \label{eq:CartesianCompleteGraphOptimalAnswerW}
     \end{gather}
\end{subequations}
An obvious example for the Cartesian product of two complete graphs is the Cartesian product of $K_2$ and $K_3$, known as the Prism graph.
The optimal value of $\lambda_2(\overline{\boldsymbol{W}})$ is $5/14$ and for the optimal transition probabilities we have $\boldsymbol{P}_{(\eta_1,\eta_2) \rightarrow (\eta_1 \pm 1,\eta_2)} = 3/7$ and $\boldsymbol{P}_{(\eta_1,\eta_2) \rightarrow (\eta_1,\eta_2 \pm 1)} = 2/7$.

\subsubsection{Complete Graph}
A complete graph with $N$ vertices is a topology where all vertices are connected to each other.
Due to the symmetry of the topology all edges have the same transition probability $(P)$, and its Laplacian operator $(\boldsymbol{L}(q))$ can be written as %
$\boldsymbol{L}(q) = (N-1)\cdot q \cdot \boldsymbol{I}  -  q \cdot ( \boldsymbol{J} - \boldsymbol{I} )  =  N \cdot q \cdot \boldsymbol{I} - q \cdot \boldsymbol{J}$,
where $\boldsymbol{J}$ is a square matrix with all elements equal to one and $q$ can be written as $q = p/N$ in terms of the transition probability $P$.
The eigenvalues of the Laplacian matrix for a complete graph are %
$\lambda_1 = 0$, $\lambda_{2}= \lambda_{3} = \cdots = \lambda_{N} = N \cdot q$.
Considering the constraint on the summation of the transition probabilities in the FCGA problem (\ref{eq:ClassicalGossipSDP}), for the complete graph topology, the optimal value of the second largest eigenvalue of the Gossip operator $(\overline{\boldsymbol{W}})$ is $\lambda_2(\overline{\boldsymbol{W}})  =  (N-2)/(N-1)$ and the transition probabilities is $P = 1 / (N-1)$.

\subsubsection{Cycle Topology}
In this topology $N$ vertices are connected in form of a cycle.
A cycle graph with four vertices in depicted in figure \ref{fig:N4Graphs}(d).
Cycle topology is edge transitive and thus the optimal transition probability $(P)$ on all edges is the same.
The optimal value of $\lambda_2(\overline{\boldsymbol{W}})$ is $\lambda_2(\overline{\boldsymbol{W}})  = ( N-  (1-\cos{(2\pi/N)})   )  / N   $ and the optimal transition probability is $P = 1/2$.

\subsubsection{Wheel Topology}

In this topology, all $n$ vertices of a circle are connected to one central vertex.
This topology has $N=n+1$ vertices.
Due to symmetry of this topology the transition probabilities can be categorized in three groups.
First and second groups are the probabilities on the edges connected to the central vertex.
These probabilities are denoted by $\boldsymbol{P}_{0,1}$ and $\boldsymbol{P}_{1,0}$.
Third group is the probability on the edges connecting the $n$ vertices in the circle to each other.
This probability is denoted by $\boldsymbol{P}_{1,1}$.

For $n < 6$, the optimal transition probabilities are %as below,
$\boldsymbol{P}_{1,1}  = (n+1)/(2\left( n+2(1-\cos{2\pi/n}) \right))$,
$\boldsymbol{P}_{1,0} = (1-2\cos{2\pi/n})/(n + 2(1-\cos{2\pi/n}))$,
$\boldsymbol{P}_{0,1}  =  1/n$,
and the optimal value of the second largest eigenvalue of the gossip operator is $(\lambda_2(\overline{\boldsymbol{W}}))  =  (  n^2 + (n-1)(1-\cos{2\pi/n})  )  /  (  n^2 + 2n(1-cos{2\pi/n})  ) $
Note that the results above for $n < 6$, are obtained for both uniform and non-uniform clock distribution.
For $n \geq 6$, the uniform clock distribution results in a optimal answer different than that of the non-uniform clock distribution where the result of obtained from uniform clock distribution is suboptimal.
For $n \geq 6$ and uniform clock distribution, the optimal transition probabilities are as below,
\begin{equation}
    \label{eq:GossipTransitionProbability1174}
    \begin{gathered}
          \boldsymbol{P}_{1,1}  = 1/2, \quad
          \boldsymbol{P}_{1,0} = 0, \quad
          \boldsymbol{P}_{0,1}  =  1/n  .
    \end{gathered}
\end{equation}
and the optimal value of the second largest eigenvalue of the gossip operator % $(\lambda_2(\overline{\boldsymbol{W}}))$ is equal to
is equal to $ \lambda_2(\overline{\boldsymbol{W}}) = (2n-1)/2n $
For $n \geq 6$ and non-uniform clock distribution, the optimal transition probabilities are as in (\ref{eq:GossipTransitionProbability1174}),
and the optimal value of the second largest eigenvalue of the gossip operator % $(\lambda_2(\overline{\boldsymbol{W}}))$ is equal to
is equal to $ \lambda_2(\overline{\boldsymbol{W}}) = (  n^2 + (n-1)(1-\cos{2\pi/n})  )  /  (  n^2 + 2n(1-cos{2\pi/n})  ) $,
and the clock distribution probabilities are
$\boldsymbol{P}_{0}  =  ( 2(1-\cos{2\pi/n}) ) / ( n+2(1-\cos{2\pi/n}) )$,
$\boldsymbol{P}_{1}  =  1 / (n+2(1-\cos{2\pi/n}))$.
\subsection{Optimization with Uniform Clock Distribution}
\label{sec:OptimizationSection}

In this subsection we address the optimization of the second largest eigenvalue of the FCGA problem (\ref{eq:ClassicalGossipSDP}) with the assumption that the clock distribution of vertices is uniform (i.e. $\boldsymbol{P}_i = 1/N$).
The complete solution procedure is provided only for the symmetric star topology.
For the rest of topologies we have provided the final results i.e. the optimal transition probabilities and the optimal value of the second largest eigenvalue of the gossip operator $(\lambda_2 (\overline{\boldsymbol{W}}))$.

\subsubsection{Symmetric Star Topology}
\label{sec:SymmetricStar}

\begin{figure}
  \centering
     \includegraphics[width=160mm]{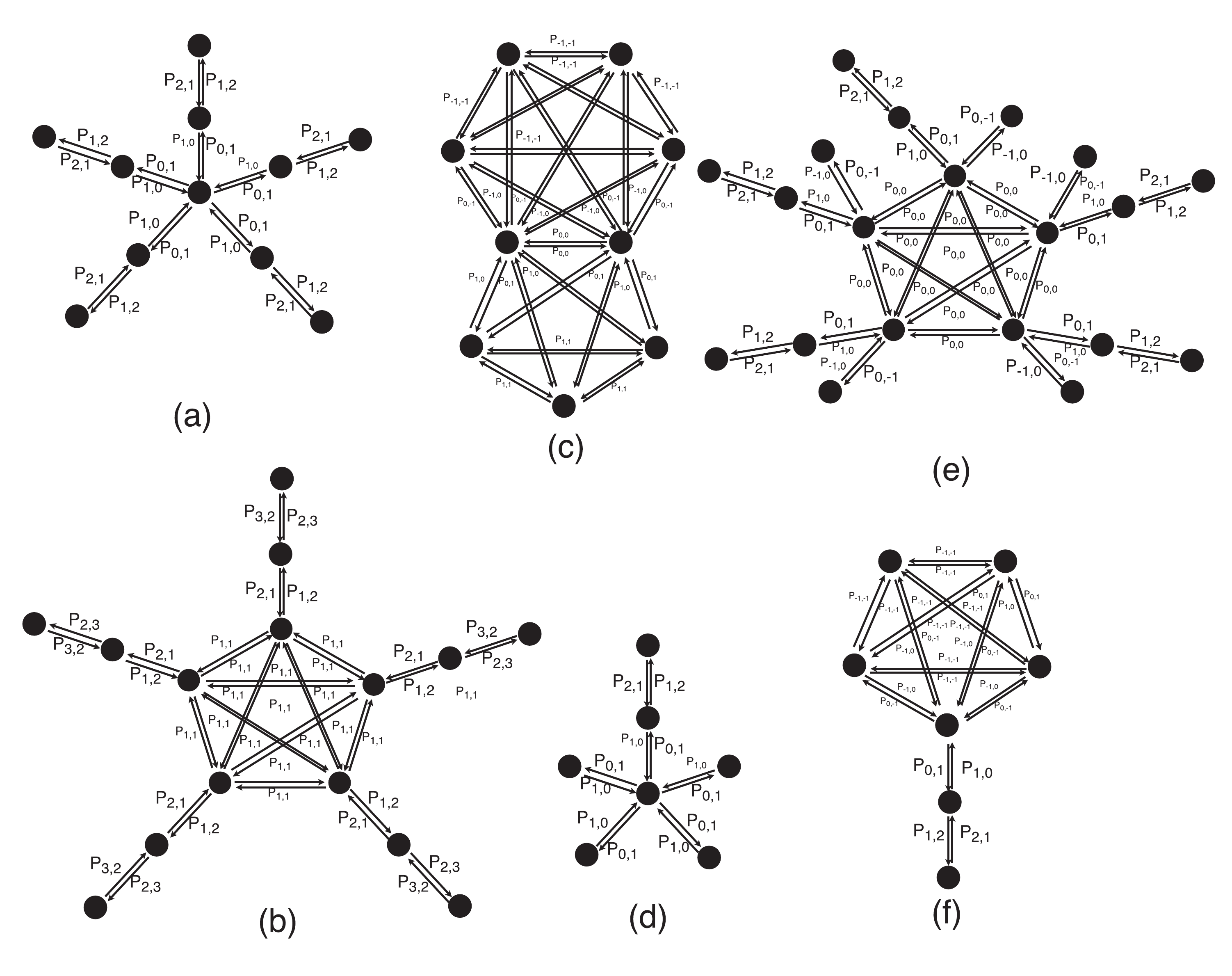}
  \caption{(a) A symmetric star graph with $n=5$ branches of length $k=2$.   (b) Complete-Cored Symmetric star topology with $n=5$ branches of length $k=2$.   (c) The two coupled complete graphs topology with parameters $N_1 = 4$, $N_2 = 2$ and $N_3 = 3$. (d) A palm graph with parameters $n=4$ and $k=2$. (e) The graph of CCS star graph with two types of branches with $n=5$ branches of length $k_1=2$ and $k_2=1$.   (f) The Lollipop graph with parameters $n=4$ and $k=2$.}
  \label{fig:GossipFigureSDP}
\end{figure}

In symmetric star topology $n$ path branches of length $k$ (each with $k$ vertices) are connected to a central vertex.
A symmetric star graph with parameters $k=2$ and $n=5$ is depicted in Figure \ref{fig:GossipFigureSDP}(a).
We model the symmeitric star topology by graph $\mathscr{G}=(\mathscr{V},\mathscr{E})$ with vertex set $\mathscr{V}=\{(0,0)\}\cup\{(i,j)|i=1,\ldots,n; j=1,\ldots,k\}$ and edge set $\mathscr{E}$.
This graph has $N = 1+k\cdot n$ vertices in total.
We associate with each vertex $(i,j)$ a column vector $e_{i,j}=e_{i}^{'} \otimes e_{j}^{''}$, for $\{i,j\}=\{0,0\}\cup\{i=1,\ldots,n;   j=1,\ldots,k\}$.
$e_{i}^{'}$ and $e_j^{''}$ are column vectors with $n+1$ and $k+1$ elements, respectively, where their $(i+1)$-th and $(j+1)$-th elements are equal to one, respectively and the rest is zero.
The gossip operator $\overline{\boldsymbol{W}}$%(\boldsymbol{P})$ % in (\ref{eq:QuantumGossipConvexFormulation})
for the symmetric star graph can be written as below,
\begin{equation}
    \nonumber
    \begin{aligned}
        \overline{\boldsymbol{W}}% _{Q}(\boldsymbol{P})
        = \boldsymbol{I} &- \sum\nolimits_{j=1}^{k-1} { q_{j,j+1} \cdot { \sum\nolimits_{i=1}^{n} (\boldsymbol{e}_{i,j} - \boldsymbol{e}_{i,j+1}) \times ( \boldsymbol{e}_{i,j} - \boldsymbol{e}_{i,j+1})^{T} } } \\
        &- q_{0,1}\cdot { \sum\limits_{i=1}^{n}  {  (\boldsymbol{e}_{0,0} - \boldsymbol{e}_{i,1}) \times (\boldsymbol{e}_{0,0} - \boldsymbol{e}_{i,1})^{T}  }  }.
    \end{aligned}
\end{equation}
Here $q_{i,j}$ is defined as $q_{i,j}=\left(\boldsymbol{P}_{i,j}+\boldsymbol{P}_{j,i} \right)/2N$  and $q_{i,j}=q_{j,i}$.
Note that $\boldsymbol{e}_{i,j} - \boldsymbol{e}_{i^{'},j^{'}}$ is associated with the edges connecting vertex $(i,j)$ to vertex $(i^{'},j^{'})$.
Due to symmetry of graph, the probabilities over edges with same distance from the central vertex are the same.
Therefore, we have only $2k-2$ variables, namely $\boldsymbol{P}_{i,i+1}$,for  $i=1,\ldots,k-1,$ and $\boldsymbol{P}_{i+1,i},$ for  $i=0,\ldots,k-2$.
Note that $\boldsymbol{P}_{0,1}=1/n$, due to symmetry of graph and $\boldsymbol{P}_{k,k-1}=1$, since $(i,k)$-th vertex for $i=1,\ldots,n$ is only connected to $(i,k-1)$-th vertex.

Due to the symmetry of symmetric star graph towards its central vertex, we can state that the automorphism group of symmetric star graph $Aut(\mathscr{G})$ is isomorphic to permutation of branches.
The orbits of $Aut(\mathscr{G})$ acting on the vertices are $\{(1,j),(2,j),\ldots,(n,j)|  j=1,\ldots,k\}$,
In order to apply stratification method, we define the new basis as below
\begin{equation}
    \begin{gathered}
        \nonumber \boldsymbol{\varphi}_{\mu,j} =
        \begin{cases}
            &\frac{1}{\sqrt{n}} \sum\limits_{i=1}^{n}{\omega^{i,\mu}\cdot \boldsymbol{e}_{i,j}}  \quad  \text{for}  \quad  \begin{array}{c} { \mu=0,\ldots,n-1 } \\ { j=1,\ldots,k } \end{array}\\
            &\boldsymbol{e}_{0,0} \quad \text{for} \quad i=\mu=0
        \end{cases}
    \end{gathered}
\end{equation}
with $\omega = e^{j\frac{2\pi}{n}}$.
On new basis $\left(\boldsymbol{\varphi}_{\mu,j}\right)$, the gossip operator $\overline{\boldsymbol{W}}$ %_{Q}(\boldsymbol{P})$
converts into smaller block diagonal matrices, namely $\overline{\boldsymbol{W}}_{0}$ and $\overline{\boldsymbol{W}}_{1}$ as below,
\begin{equation}
\begin{gathered}
    \nonumber
    \overline{\boldsymbol{W}}_{0} =
    \small{\left[ \begin{array}{cccccc}
    {1-n\cdot q_{0,1}}       &{\sqrt{n}\cdot q_{0,1}}  &{0}                    &{\cdots}      &{0} \\
    {\sqrt{n}\cdot q_{0,1}}  &{1-q_{0,1}-q_{1,2}}      &{q_{1,2}}              &{\cdots}      &{\vdots}  \\
    {0}                      &{q_{1,2}}                &{1-q_{1,2}-q_{2,3}}    &{\ddots}      &{0} \\
    {\vdots}                 &{\vdots}                 &{\ddots}               &{\ddots}      &{q_{k-1,k}} \\
    {0}                      &{\cdots}                 &{0}                    &{q_{k-1,k}}   &{1-q_{k-1,k}}  \end{array} \right]},
\end{gathered}
\end{equation}
\begin{equation}
\begin{gathered}
    \nonumber
    \overline{\boldsymbol{W}}_1 =
    \small{\left[ \begin{array}{ccccc}
    {1-q_{0,1}-q_{1,2}}     & {q_{1,2}}             & {0}                   &{\cdots}                   &{0} \\
    {q_{1,2}}               & {1-q_{1,2}-q_{2,3}}   & {q_{2,3}}             & {\ddots}                  &{\vdots} \\
    {0}                     & {q_{2,3}}             & {1-q_{2,3}-q_{3,4}}   & {\ddots}                  &{0} \\
    {\vdots}                & {\ddots}              & {\ddots}              & {\ddots}                  & {q_{k-1,k}} \\
    {0}                     & {\cdots}              & {0}                   & {q_{k-1,k}}               & {1-q_{k-1,k}}  \end{array} \right]},
\end{gathered}
\end{equation}
Note that $\overline{\boldsymbol{W}}_{1}$ is a submatrix of $\overline{\boldsymbol{W}}_{0}$.
According to Cauchy Interlacing theorem \cite{CauchyInterlacingRef}, for the eigenvalues of $\overline{\boldsymbol{W}}_1$ and $\overline{\boldsymbol{W}}_0$ we have $\lambda_{k+1}(\overline{\boldsymbol{W}}_0)\leq \lambda_k(\overline{\boldsymbol{W}}_1)\leq \cdots \leq \lambda_2(\overline{\boldsymbol{W}}_1) \leq \lambda_2(\overline{\boldsymbol{W}}_0) \leq \lambda_1(\overline{\boldsymbol{W}}_1) \leq \lambda_1(\overline{\boldsymbol{W}}_0)=1$.
From these relations we can conclude that the second largest eigenvalue of $\overline{\boldsymbol{W}}$ %
is included in $\overline{\boldsymbol{W}}_1$ and thus the original optimization problem reduces to the following minimization problem,
\begin{equation}
    \nonumber
    \begin{aligned}
        \min \quad &s \\
        s.t. \quad &\overline{\boldsymbol{W}}_1 \preceq s\mathbf{I},\\
        &1-P_{i,i-1} - P_{i,i+1} \geq 0, \quad \text{for} \quad i=1,\ldots,k.
    \end{aligned}
\end{equation}
$m$ is the number of edges (starting from the central vertex) which their transition probabilities have not reached the boundaries of feasible region.
In other words they can take values between one and zero.
The transition probabilities over these edges are $\boldsymbol{P}_{i,i-1}$ and $\boldsymbol{P}_{i,i+1}$ for $i=1, \ldots, m$.
For the rest of the edges, the transition probabilities on the edges towards the center of graph are one, i.e. $\boldsymbol{P}_{i,i-1} = 1$   for $i=m+1,\ldots,k-1$,
and the transition probabilities in the opposite direction are zero, i.e. $\boldsymbol{P}_{i,i+1} = 0$ for $i=m+1,\ldots,k-1$.
Note that we have $2m$ variables in the optimization problem.
The values of variables $q_{i,i+1}$ can be written as below,
\begin{subequations}
\label{eq:Gossip42}
    \begin{align}
            &q_{0,1}= \frac{ 1/n + \boldsymbol{P}_{1,0} } {2N}, \label{eq:Gossip42a} \\
            &q_{i,i+1} = \frac{ \boldsymbol{P}_{i,i+1} + \boldsymbol{P}_{i+1,i} }  {2N}, \quad \text{for} \quad i=1,\ldots, m-1 \label{eq:Gossip42b}\\
            &q_{m,m+1}=\frac{\boldsymbol{P}_{m,m+1}+1} {2N}  \label{eq:Gossip42c} \\
            &q_{i,i+1} = \frac{1}{2N} \quad \text{for} \quad i=m+1,\ldots,k-1. \label{eq:Gossip42d}
    \end{align}
\end{subequations}
The problem parameters $(\boldsymbol{F}_{i}, \boldsymbol{c})$ of the standard semidefinite programming {\cite[Appendix~B]{SaberQConsensusContinuous}} are as following
\begin{equation}
    \nonumber
    \begin{gathered}
    \boldsymbol{F}_{1,0} =
    \left[ \begin{array}{cc}
    {  \frac{1}{2N} \boldsymbol{e}_{1}^{'}  \times  {\boldsymbol{e}_{1}^{'}}^{T}  } & {  0  }  \\
    {  0  } & {  -\boldsymbol{e}_{1}^{''} \times {\boldsymbol{e}_{1}^{''}}^{T}  }     \end{array} \right],
    \end{gathered}
\end{equation}
\begin{equation}
    \nonumber
    \begin{gathered}
    \boldsymbol{F}_{i,i+1} =  %
    \left[ \begin{array}{cc}
    {  \frac{1}{2N}( \boldsymbol{e}_{i}^{'} - \boldsymbol{e}_{i+1}^{'} ) \times ( \boldsymbol{e}_{i}^{'} - \boldsymbol{e}_{i+1}^{'} )^{T}  } & {  0  }  \\
    {  0  } & {  -\boldsymbol{e}_{i}^{''} \times {\boldsymbol{e}_{i}^{''}}^{T}  }     \end{array} \right], \\
    \text{for} \quad i=1,\ldots,m,
    \end{gathered}
\end{equation}
\begin{equation}
    \nonumber
    \begin{gathered}
    \boldsymbol{F}_{i+1,i} =  % \hspace{330pt} \\
    \left[ \begin{array}{cc}
    {  \frac{1}{2N}( \boldsymbol{e}_{i}^{'} - \boldsymbol{e}_{i+1}^{'} ) \times ( \boldsymbol{e}_{i}^{'} - \boldsymbol{e}_{i+1}^{'} )^{T}  } & {  0  }  \\
    {  0  } & {  -\boldsymbol{e}_{i+1}^{''} \times {\boldsymbol{e}_{i+1}^{''}}^{T}  }     \end{array} \right], \\
    \text{for} \quad i=1,\ldots,m-1,
    \end{gathered}
\end{equation}
\begin{equation}
    \nonumber
    \begin{gathered}
    \boldsymbol{F}_{s} =
    \left[ \begin{array}{cc}
    {  \boldsymbol{I}_{k \times k}  } & {  0  }  \\
    {  0  } & {  0  }     \end{array} \right],
    \quad
    \boldsymbol{F}_{0} =
    \left[ \begin{array}{cc}
    {  -\boldsymbol{I}_{k \times k}  } & {  0  }  \\
    {  0  } & {  \boldsymbol{I}_{m \times m}  }     \end{array} \right].
    \end{gathered}
\end{equation}
Here $\boldsymbol{e}_{i}^{'}$ for $i=1,\ldots,k$ is a column vector with $k$ elements, where its $i$-th element is equal to one and the rest is zero.
$\mathbf{e}_{i}^{''}$ for $i=1,\ldots,m$ is a column vector with $m$ elements, where its $i$-th element is equal to one and the rest is zero.
$\mathbf{c}$ is a vector with length $1+2m$ with $1$ at index $1+2m$ and zero elsewhere.
Minimization variable $(\mathbf{x})$ can be defined as $\boldsymbol{x}$ $=$ $[\boldsymbol{P}_{1,0}$, $\ldots$, $\boldsymbol{P}_{m,m-1}$, $\boldsymbol{P}_{1,2}$, $\ldots$,  $\boldsymbol{P}_{m,m+1}$, $s ]^T$.
For the dual problem we choose the dual variable $\boldsymbol{Z}$ as $\boldsymbol{Z}=\boldsymbol{z}\times \boldsymbol{z}^{T}$, where $\boldsymbol{z}$ is a column vector defined as
$   \boldsymbol{z}  =  \left[ z_{1}, \cdots, z_{k}, \xi_{1}, \cdots, \xi_{m} \right]^T   $.
From the constraints of dual problem $(tr[\boldsymbol{F}_{i}\times \boldsymbol{Z}]=\boldsymbol{c}_{i})$ we obtain the following relations,
$( \boldsymbol{z}_{1}^{2}) /  2N = \xi_{1}^{2}$, and
$\left( ( \boldsymbol{z}_{i}  -  \boldsymbol{z}_{i+1} )^{2} \right) / 2N = \xi_{i}^{2}$, for $i = 1, \ldots, m-1$, and
$\left( ( \boldsymbol{z}_{i}  -  \boldsymbol{z}_{i+1} )^{2} \right) / 2N = \xi_{i+1}^{2}$, for $i = 1, \ldots, m-1$, and
$\left( ( \boldsymbol{z}_{m}  -  \boldsymbol{z}_{m+1} )^{2} \right) / 2N = \xi_{m}^{2}$ and
\begin{equation}
    \label{eq:Gossip111}
    \begin{gathered}
    \xi_{j} (1-\boldsymbol{P}_{j,j-1} - \boldsymbol{P}_{j,j+1} )=0, \quad \text{for}  j=1,\ldots,m.
    \end{gathered}
\end{equation}
From these equations, we can conclude the following relations,
\begin{equation}
    \label{eq:Gossip12}
    \begin{gathered}
    z_{i} = i \cdot \boldsymbol{z}_{1}, \quad \text{for} \quad  i=1,\ldots,m+1.
    \end{gathered}
\end{equation}
From complementary slackness condition ( $\boldsymbol{F}(\hat{\boldsymbol{x}})  \times  \hat{\boldsymbol{Z}}  =  \hat{\boldsymbol{Z}}  \times  \boldsymbol{F}(\hat{\boldsymbol{x}}) = 0$ where $\hat{\boldsymbol{x}}$ and $\hat{\boldsymbol{Z}}$ are the optimal primal dual feasible points) we have $\left( s\cdot \boldsymbol{I} - \overline{\boldsymbol{W}}_1 \right)\times \boldsymbol{Z} = 0   \Rightarrow   \left(s\cdot \boldsymbol{I} - \overline{\boldsymbol{W}}_1 \right)\times \boldsymbol{z}=0$.
This in turn results in the following relations for optimal values of primal feasible point $(\boldsymbol{x})$ and dual feasible point $(\boldsymbol{Z})$,
\begin{subequations}
\label{eq:Gossip13}
    \begin{gather}
            (s-1)\cdot \boldsymbol{z}_1 + (q_{0,1} + q_{1,2})\cdot \boldsymbol{z}_1 - q_{1,2}\cdot \boldsymbol{z}_2 = 0, \label{eq:Gossip13a} \\
            (s-1)\cdot \boldsymbol{z}_i - q_{i-1,i}\cdot \boldsymbol{z}_{i-1}    +  (q_{i-1,i}+q_{i,i+1} )\cdot \boldsymbol{z}_i - q_{i,i+1} \cdot \boldsymbol{z}_{i+1}=0,    \quad \text{for} \quad i=2, \ldots, m-1, \label{eq:Gossip13b} \\
            (s-1)\cdot \boldsymbol{z}_{m} - q_{m-1,m}\cdot \boldsymbol{z}_{m-1} ( q_{m-1,m} + q_{m,m+1} ) \cdot \boldsymbol{z}_{m} - q_{m,m+1} \cdot \boldsymbol{z}_{m+1} = 0, \label{eq:Gossip13c} \\
            (s-1)\cdot \boldsymbol{z}_{m+1} - q_{m,m+1}\cdot \boldsymbol{z}_{m} + ( q_{m,m+1} + \frac{1}{2N} ) \cdot \boldsymbol{z}_{m+1} - \frac{1}{2N} \cdot \boldsymbol{z}_{m+2} = 0, \label{eq:Gossip13d} \\
            (s-1)\cdot \boldsymbol{z}_{i} - \frac{1}{2N} \cdot \boldsymbol{z}_{i-1} +  \frac{2}{2N}  \cdot \boldsymbol{z}_{i} - \frac{1}{2N} \cdot \boldsymbol{z}_{i+1} = 0,  \; \text{for} \; i = m+2, \ldots, k-1, \label{eq:Gossip13e} \\
            \normalsize{ (s-1)\cdot \boldsymbol{z}_k - \frac{1}{2N} \cdot \boldsymbol{z}_{k-1} + \frac{1}{2N} \cdot \boldsymbol{z}_k = 0, } \label{eq:Gossip13f}
    \end{gather}
\end{subequations}
By substituting (\ref{eq:Gossip12}) in (\ref{eq:Gossip13a}) to (\ref{eq:Gossip13d}) we get the following equations,
\begin{subequations}
\label{eq:Gossip14}
    \begin{gather}
        (s-1) + q_{0,1} - q_{1,2}  = 0,  \label{eq:Gossip14a}\\
        (s-1) \cdot i + q_{i-1,i} - q_{i,i+1} = 0,    \quad \text{for} \quad  i=2,\ldots,m-1     \label{eq:Gossip14b}   \\
        (s-1) \cdot m + q_{m-1,m} - q_{m,m+1} = 0,   \label{eq:Gossip14c}\\
        \left( (s-1) + \frac{q_{m,m+1}}{m+1} + \frac{1}{2N} \right) \cdot \boldsymbol{z}_{m+1} - \frac{1}{2N} \cdot \boldsymbol{z}_{m+2} = 0,   \label{eq:Gossip14d}
    \end{gather}
\end{subequations}
Defining $\widetilde{q}_{i,j}=(\boldsymbol{P}_{i,j} + \boldsymbol{P}_{j,i} )=2N\cdot q_{i,j}$, and $X=2N(s-1)$, the equations (\ref{eq:Gossip14a}) to (\ref{eq:Gossip14c}) can be written as % below,
$i\cdot X  +  \widetilde{q}_{i-1,i}  -  \widetilde{q}_{i,i+1}=0$  for  $i=1,\ldots,m$,
or equivalently,
\begin{equation}
    \label{eq:Gossip15}
    \begin{gathered}
        \widetilde{q}_{i,i+1} = \widetilde{q}_{0,1}+\frac{i\cdot (i+1)}{2}\cdot X, \quad \text{for} \quad  i=1,\ldots,m.
    \end{gathered}
\end{equation}
In order to satisfy (\ref{eq:Gossip111}) either $\xi_{j}$ should be set to zero or $(1-\boldsymbol{P}_{j,j-1}$ $-$ $\boldsymbol{P}_{j,j+1})=0$, where the former is not acceptable, since setting $\xi_{j}=0$ for $j=1,\ldots,m$ will result in all elements of the vector $\boldsymbol{z}$ to be zero.
Thus the following can be concluded from (\ref{eq:Gossip111})
\begin{equation}
    \label{eq:Gossip16}
    \begin{gathered}
        1 - \boldsymbol{P}_{j,j-1} - \boldsymbol{P}_{j,j+1}=0, \quad \text{for} \quad  j=1,\ldots,m,
    \end{gathered}
\end{equation}
Substituting $\widetilde{q}_{i,i+1}=\left(\boldsymbol{P}_{i,i+1} + \boldsymbol{P}_{i+1,i}\right)$ and (\ref{eq:Gossip16}) in (\ref{eq:Gossip15}) we get
\begin{equation}
    \label{eq:Gossip17}
    \begin{gathered}
        \boldsymbol{P}_{i+1,i}  =  i \cdot \left( \frac{1}{n} - 1 \right)  +  (i+1) \cdot \boldsymbol{P}_{1,0} + \frac{i(i+1)(i+2)}{6}\cdot X,
        \quad
        \text{for} \quad i=1, \ldots, m.
    \end{gathered}
\end{equation}
From equations (\ref{eq:Gossip15}) and (\ref{eq:Gossip16}) for $i=m$, we have
\begin{equation}
    \label{eq:Gossip18}
    \begin{gathered}
        \boldsymbol{P}_{m,m-1}  =  2  -  \frac{1}{n} - \boldsymbol{P}_{1,0}  -  \frac{m \cdot (m+1)}{2} \cdot X.
    \end{gathered}
\end{equation}
On the other hand, from (\ref{eq:Gossip17}) for $i=m-1$ we have
\begin{equation}
    \label{eq:Gossip19}
    \begin{gathered}
        \boldsymbol{P}_{m,m-1}  =  (m-1) \cdot \left( 1 -  \frac{1}{n} \right) +  %
        m \cdot \boldsymbol{P}_{1,0}  +  \frac{m \cdot (m-1) \cdot (m+1)}{6} \cdot X %.
    \end{gathered}
\end{equation}
Comparing the equations (\ref{eq:Gossip18}) and (\ref{eq:Gossip19}) we can conclude the following
\begin{equation}
    \label{eq:Gossip20}
    \begin{gathered}
        \boldsymbol{P}_{1,0}  =  1  -  \frac{m} { n \cdot (m+1) }  -  \frac{ m(m+2) } {6} \cdot X,
    \end{gathered}
\end{equation}
and thus for $\widetilde{q}_{m,m+1}$ we have
\begin{equation}
    \label{eq:Gossip21}
    \begin{gathered}
        \widetilde{q}_{m,m+1}  =  \frac{1} { n \cdot (m+1)}  +   1   +    \frac{2 m^2 + m} {6}  \cdot X.
    \end{gathered}
\end{equation}
Substituting (\ref{eq:Gossip21}) in (\ref{eq:Gossip14d}) we obtain the following
\begin{equation}
    \label{eq:Gossip22}
    \begin{gathered}
        \left( 6 \left(  1  +  n  (m+1)  (m+2)  \right)  +  n  (m+1)  %
        ( 2 m^2 + 7m +6 )  X  \right) \cdot  \boldsymbol{z}_{m+1}    -    6 n (m+1)^2  \cdot  \boldsymbol{z}_{m+2}  =  0.
    \end{gathered}
\end{equation}
From equations (\ref{eq:Gossip13e}) and (\ref{eq:Gossip13f}) we obtain the following inductive equations
\begin{subequations}
\label{eq:Gossip23}
    \begin{gather}
        X \cdot \boldsymbol{z}_{i} - \boldsymbol{z}_{i-1} + 2\boldsymbol{z}_{i} - \boldsymbol{z}_{i+1}  =  0,    \label{eq:Gossip23a} \\
        X \cdot \boldsymbol{z}_{m} - \boldsymbol{z}_{m-1} + \boldsymbol{z}_{m}  =  0,   \label{eq:Gossip23b}
    \end{gather}
\end{subequations}
where (\ref{eq:Gossip23a}) holds for $i = m+2,\ldots,k-1 $.
Using the recursive equations (\ref{eq:Gossip23}) we can calculate $\boldsymbol{z}_{i}$ for $i=m+2, \ldots, k$ in terms of $\boldsymbol{z}_{m}$ as below,
\begin{equation}
    \label{eq:Gossip24}
    \begin{gathered}
        \boldsymbol{z}_{k-i}  =  F_{i}(X) \cdot \boldsymbol{z}_{k}, \quad \text{for} \quad i = 1, \ldots, k-m-1
    \end{gathered}
\end{equation}
where $F_{i}(X)$ is a polynomial of order $i$ in terms of X.
By substituting  $\boldsymbol{z}_{m+1} = F_{k-m-1}(X) \cdot \boldsymbol{z}_{k}$ and $\boldsymbol{z}_{m+2} = F_{k-m-2}(X) \cdot \boldsymbol{z}_{k}$ in (\ref{eq:Gossip22}), we get the following polynomial,
\begin{equation}
    \label{eq:Gossip25}
    \begin{aligned}
        \left(  6 \left( 1 + n  (m+1)  (m+2) \right)  +  n  (m+1)  %\right. \\
        \left( 2 m^2 + 7 m + 6 \right)  X  \right)  F_{k-m-1}(X)  % \\
        -  6 n  (m+1)^2  F_{k-m-2}(X)  =  0
    \end{aligned}
\end{equation}
We refer to this polynomial as the final polynomial.
This is a polynomial of order $k-m+2$ in terms of $X$.
All roots of this polynomial are in interval $(0,1)$ where $s$ is obtained from the largest root $(Xr)$ of the final polynomial (\ref{eq:Gossip25}) according to the formula $s = ( X_r / 2N ) + 1$.
After finding the optimal value of $X$ and correspondingly $s$, the optimal value of probabilities can be obtained by substituting the optimal value of $X$ in formulas (\ref{eq:Gossip16}), (\ref{eq:Gossip17}), (\ref{eq:Gossip20}).

The optimal results presented above are obtained for a given $m$.
The correct value of $m$ is the largest value where the resultant optimal probabilities and the second largest eigenvalue $(s)$ are within the feasible region (i.e. between zero and one).
To find this value of $m$, first one should set the value of $m$ to its minimum value i.e. zero and then calculate the optimal probabilities and $s$.
If the obtained values are acceptable and they are within the feasible region (i.e. between zero and one) then it should increase the value of $m$ and redo the calculation.
This process continues until the smallest value of $m$ that results in optimal probabilities outside of feasible region is obtained.
In Algorithm \ref{mFindingAlgorithm} we have provided the pseudo code for this process.
\begin{algorithm}
\caption{   \small{Finding the optimal value of $m$}   }
\label{mFindingAlgorithm}
\small{\textbf{Input:} The final polynomial for determining $s$, i.e. formula (\ref{eq:Gossip25}) }  \\*
\text{\hspace{32pt}} \small{ Optimal Probabilities in terms of $s$, i.e.formulas (\ref{eq:Gossip16}), (\ref{eq:Gossip17}) and (\ref{eq:Gossip18}) }  \\
\small{\textbf{Output:} Optimal value of $m$}

\begin{algorithmic}[1]
\State $i \leftarrow 0$      \Comment{Starting from the minimum possible value (i.e. zero)}
\State Set $m$ equal to $i$ and calculate the optimal value of $s$ and probabilities using final polynomial (\ref{eq:Gossip25}) and (\ref{eq:Gossip16}), (\ref{eq:Gossip17}), (\ref{eq:Gossip20})   \Comment{Calculating the optimal answers} \label{marker724}
\If{ \hspace{-2pt} the obtained transition probabilities and $s$ are inside the feasible region  \hspace{7pt} } \Comment{Checking if $m$ is feasible} \\ {  \hspace{20pt}  Increase the value of $i$ by one and  }
\hspace{0pt} \Goto{marker724}
\Else\Comment{Obtained $m$ is the smallest value of $m$ in feasible region } \\
\hspace{20pt} Set the optimal value of $m$ to $i-1$ and exit
\EndIf
\end{algorithmic}
\end{algorithm}
For the rest of topologies presented in this section the main procedure for finding the optimal value of probabilities and the second largest eigenvalue $(s)$ is more or less the same as Algorithm \ref{mFindingAlgorithm}.
The main difference is on the final polynomial and the formulas determining the optimal probabilities in terms of the second largest eigenvalue $(s)$ or equivalently $X$.

In Tables \ref{tab:GossipTableSS1} and \ref{tab:GossipTableSS2} we have provided the optimal value of $m$ and $s$ (second largest eigenvalue) for different symmetric star topologies.
The results in these tables are presented in order to analyse the optimal value of $m$ and $s$ in terms of topological parameters of symmetric star.
\begin{table}
\centering
    \caption{The optimal value of $m$ in terms of number $(n)$ and length $(k)$ of branches for a symmetric star topology .}
    \vspace{-2pt}
    \label{tab:GossipTableSS1}
    \begin{tabular}{|c|c|c|c|c|c|c|} \hline
    {} & {$n=3$} & {$n=4$} & {$n=5$} & {$n=6$} & {$n=7$} & {$n=8$} \\ \hline
    {$k=2$} & {$0$} & {$0$} & {$0$} & {$0$} & {$0$} & {$0$} \\ \hline
    {$k=3$} & {$1$} & {$1$} & {$0$} & {$0$} & {$0$} & {$0$} \\ \hline
    {$k=4$} & {$1$} & {$1$} & {$1$} & {$1$} & {$1$} & {$0$} \\ \hline
    {$k=5$} & {$1$} & {$1$} & {$1$} & {$1$} & {$1$} & {$1$} \\ \hline
    {$k=6$} & {$2$} & {$1$} & {$1$} & {$1$} & {$1$} & {$1$} \\ \hline
    {$k=7$} & {$2$} & {$2$} & {$1$} & {$1$} & {$1$} & {$1$} \\ \hline
    {$k=8$} & {$2$} & {$2$} & {$2$} & {$1$} & {$1$} & {$1$} \\ \hline
    {$k=9$} & {$2$} & {$2$} & {$2$} & {$2$} & {$2$} & {$1$} \\ \hline
    {$k=10$} & {$3$} & {$2$} & {$2$} & {$2$} & {$2$} & {$2$} \\ \hline
    \end{tabular}
    \vspace{-12pt}
\end{table}
\begin{table}
\centering
    \caption{The optimal value of $s$ (second largest eigenvalue of the gossip operator $\overline{\boldsymbol{W}}$ in terms of number $(n)$ and length $(k)$ of branches for a symmetric star topology.}
    \vspace{-2pt}
    \label{tab:GossipTableSS2}
    \begin{tabular}{|c|c|c|c|c|c|c|} \hline
    {\hspace{-3pt}$k$\hspace{-2pt}} & {\hspace{-6pt}$n=3$\hspace{-7pt}} & {\hspace{-6pt} $n=4$ \hspace{-7pt}} & {\hspace{-6pt} $n=5$ \hspace{-7pt}} & {\hspace{-6pt} $n=6$ \hspace{-7pt}} & {\hspace{-6pt} $n=7$ \hspace{-7pt}} & {\hspace{-6pt} $n=8$ \hspace{-7pt}} \\ \hline
    {\hspace{-3pt}$2$\hspace{-2pt}} & {\hspace{-6pt}$0.971428$\hspace{-7pt}} & {\hspace{-6pt} $0.97863247$ \hspace{-7pt}} & {\hspace{-6pt} $0.98295454$ \hspace{-7pt}} & {\hspace{-6pt} $0.98582995$ \hspace{-7pt}} & {\hspace{-6pt} $0.987878$ \hspace{-7pt}} & {\hspace{-6pt} $0.9894117$ \hspace{-7pt}} \\ \hline
    {\hspace{-3pt}$3$\hspace{-2pt}} & {\hspace{-6pt} $0.988548$ \hspace{-7pt}} & {\hspace{-6pt} $0.99146614$ \hspace{-7pt}} & {\hspace{-6pt} $0.99320128$ \hspace{-7pt}} & {\hspace{-6pt} $0.9942928$ \hspace{-7pt}} & {\hspace{-6pt} $0.995122$ \hspace{-7pt}} & {\hspace{-6pt} $0.995741$ \hspace{-7pt}} \\ \hline
    {\hspace{-3pt}$4$\hspace{-2pt}} & {\hspace{-6pt} $0.994781$ \hspace{-7pt}} & {\hspace{-6pt} $0.996114$ \hspace{-7pt}} & {\hspace{-6pt} $0.996906$ \hspace{-7pt}} & {\hspace{-6pt} $0.997430$ \hspace{-7pt}} & {\hspace{-6pt} $0.9978028$ \hspace{-7pt}} & {\hspace{-6pt} $0.9980817$ \hspace{-7pt}} \\ \hline	
    {\hspace{-3pt}$5$\hspace{-2pt}} & {\hspace{-6pt} $0.997205$ \hspace{-7pt}} & {\hspace{-6pt} $0.997917$ \hspace{-7pt}} & {\hspace{-6pt} $0.998341$ \hspace{-7pt}} & {\hspace{-6pt} $0.998622$ \hspace{-7pt}} & {\hspace{-6pt} $0.998822$ \hspace{-7pt}} & {\hspace{-6pt} $0.998971$ \hspace{-7pt}} \\ \hline
    {\hspace{-3pt}$6$\hspace{-2pt}} & {\hspace{-6pt} $0.998334$ \hspace{-7pt}} & {\hspace{-6pt} $0.998758$ \hspace{-7pt}} & {\hspace{-6pt} $0.999011$ \hspace{-7pt}} & {\hspace{-6pt} $0.999178$ \hspace{-7pt}} & {\hspace{-6pt} $0.999297$ \hspace{-7pt}} & {\hspace{-6pt} $0.999386$ \hspace{-7pt}} \\ \hline
    {\hspace{-3pt}$7$\hspace{-2pt}} & {\hspace{-6pt} $0.998929$ \hspace{-7pt}} & {\hspace{-6pt} $0.999201$ \hspace{-7pt}} & {\hspace{-6pt} $0.999363$ \hspace{-7pt}} & {\hspace{-6pt} $0.999471$ \hspace{-7pt}} & {\hspace{-6pt} $0.999547$ \hspace{-7pt}} & {\hspace{-6pt} $0.99960$ \hspace{-7pt}} \\ \hline	
    {\hspace{-3pt}$8$\hspace{-2pt}} & {\hspace{-6pt} $0.999272$ \hspace{-7pt}} & {\hspace{-6pt} $0.999456$ \hspace{-7pt}} & {\hspace{-6pt} $0.999567$ \hspace{-7pt}} & {\hspace{-6pt} $0.999640$ \hspace{-7pt}} & {\hspace{-6pt} $0.999692$ \hspace{-7pt}} & {\hspace{-6pt} $0.999731$ \hspace{-7pt}} \\ \hline	
    {\hspace{-3pt}$9$\hspace{-2pt}} & {\hspace{-6pt} $0.999482$ \hspace{-7pt}} & {\hspace{-6pt} $0.999614$ \hspace{-7pt}} & {\hspace{-6pt} $0.999692$ \hspace{-7pt}} & {\hspace{-6pt} $0.999744$ \hspace{-7pt}} & {\hspace{-6pt} $0.999780$ \hspace{-7pt}} & {\hspace{-6pt} $0.999808$ \hspace{-7pt}} \\ \hline	
    {\hspace{-3pt}$10$\hspace{-2pt}} & {\hspace{-6pt} $0.999619$ \hspace{-7pt}} & {\hspace{-6pt} $0.999715$ \hspace{-7pt}} & {\hspace{-6pt} $0.999773$ \hspace{-7pt}} & {\hspace{-6pt} $0.999811$ \hspace{-7pt}} & {\hspace{-6pt} $0.999838$ \hspace{-7pt}} & {\hspace{-6pt} $0.999859$ \hspace{-7pt}} \\ \hline
    \end{tabular}
    \vspace{-16pt}
\end{table}
Based on the results presented in Table \ref{tab:GossipTableSS1}, we can state that for fixed length of branches in symmetric star topology, the optimal value of $m$ decreases as the number of branches increases and as expected the optimal value of $m$ increases by the length of branches.
This can be translated into the fact that the optimal point gets closer to the borders of feasible region as the number of branches increases.
From the results presented in Table \ref{tab:GossipTableSS2} it is obvious that by adding to both the length and the number of branches in symmetric star topology, the optimal value of second largest eigenvalue $(s)$ increases.

\subsubsection{Path Topology}
\label{sec:PathOptimizationUniform}

Due to the symmetry of path graph towards its center, we can state that the automorphism group of path graph is isomorphic to permutation of two path branches obtained from dividing the whole graph from its center.
Therefore, for a path topology with even number of vertices (i.e. $2k$ vertices), we can denote the independent transition probabilities as $\boldsymbol{P}_{i,i+1}$  for  $i=1,\ldots,k$ and $\boldsymbol{P}_{i+1,i}$ for  $i=1,\ldots,k-1$.
Similarly for a path topology with odd number of vertices (i.e. $2k+1$ vertices), the independent transition probabilities are $\boldsymbol{P}_{i,i+1}$ for$i=1,\ldots,k$ and $\boldsymbol{P}_{i+1,i}$ for $i=1,\ldots,k$.
Note that for the path topology with even and odd number of vertices, center of the topology refers to the edge $(k,k+1)$ and the vertex $(k+1)$, respectively.
Here $m$ is the number of edges (starting from center of the path topology) which their transition probabilities have not reached the boundaries of feasible region.

\subsubsection*{Path with Even Vertices}
Following the same routine performed for symmetric star topology, we get the following results for the optimal transition probabilities and second largest eigenvalue $(s)$ of the path topology with even number of vertices (i.e. $2k$ vertices),
\begin{subequations}
    \label{eq:Gossip1403}
    \begin{align}
          &\boldsymbol{P}_{i,i+1} = 1, \quad \text{for} \quad  i=1,\ldots,k-m, \\
          &\boldsymbol{P}_{i+1,i} = 0, \quad \text{for} \quad  i=1,\ldots,k-m-1,
    \end{align}
\end{subequations}
\begin{equation}
    \label{eq:Gossip1402}
    \begin{gathered}
          1-\boldsymbol{P}_{k-j+1,k-j}  -  \boldsymbol{P}_{k-j+1,k-j+2}=0, \; \text{for} \;  j=1,\ldots,m,
    \end{gathered}
\end{equation}
\begin{equation}
    \label{eq:Gossip1411}
    \begin{gathered}
          \boldsymbol{P}_{k-i,k-i+1}=(2i+1)\cdot P_{k,k+1}-i + \frac{i(i+1)(2i+1)}{12} X, \\
          \text{for} \quad  i=1,\ldots,m-1
    \end{gathered}
\end{equation}
\begin{equation}
    \label{eq:Gossip1410}
    \begin{gathered}
        \boldsymbol{P}_{k,k+1} = \frac { 12(m+1)-\left(6m^2+(m-1)m(2m-1) \right)X } {12(2m+1)}
    \end{gathered}
\end{equation}
\begin{equation}
    \label{eq:Gossip1426}
    \begin{gathered}
            \left(12m^2+24m+15+\left(4m^3+12m^2+11m+3\right)\cdot X\right)
            \cdot F_{k-m}(X)   %
            -3(2m+1)^2\cdot F_{k-m-1}(X)=0
    \end{gathered}
\end{equation}
(\ref{eq:Gossip1426}) is the final polynomial for the path topology with even number of vertices.
This is a polynomial of order $k-m+2$ in terms of X.
All roots of this polynomial are negative where $s$ is obtained from the largest root $(X_r)$ of the final polynomial (\ref{eq:Gossip1426}) according to the formula $s=(X_r/2N+1$.
After finding the optimal value of $X$ and correspondingly $s$, the optimal value of probabilities can be obtained by substituting the optimal value of $X$ in formulas (\ref{eq:Gossip1403}), (\ref{eq:Gossip1402}), (\ref{eq:Gossip1411}) and (\ref{eq:Gossip1410}).
The polynomials $F_{k-m}(X)$ and $F_{k-m-1}(X)$ are polynomials of order $k-m-1$ and $k-m-2$, respectively and they can be calculated recursively from %
equations $F_{1}(X) = 1$, $F_{2}(X) = X+1$ and $F_{i+1}(X) = (X+2) \cdot F_{i}(X) - F_{i-1}(X)$ for $i=2, \ldots, k-m-1$.
Note that $F_{i}(X)$ is a polynomial of order $i-1$ in terms of $X$.
Similar to the symmetric star topology, here for finding the optimal value of $m$, Algorithm \ref{mFindingAlgorithm} should be performed with the final polynomial given in (\ref{eq:Gossip1426}) and the optimal probabilities given in terms of $X$ in equations (\ref{eq:Gossip1403}), (\ref{eq:Gossip1402}), (\ref{eq:Gossip1411}) and (\ref{eq:Gossip1410}).

\subsubsection*{Path with Odd Vertices}
The followings are the results for the optimal transition probabilities and second largest eigenvalue $(s)$ of the path topology with odd number of vertices (i.e. $2k+1$ vertices),
\begin{subequations}
    \label{eq:Gossip1473}
    \begin{align}
          &\boldsymbol{P}_{i,i+1} = 1, \quad \text{for} \quad  i=1,\ldots,k-m, \\
          &\boldsymbol{P}_{i+1,i} = 0, \quad \text{for} \quad  i=1,\ldots,k-m-1,
    \end{align}
\end{subequations}
\begin{equation}
    \label{eq:Gossip1482}
    \begin{gathered}
          1-\boldsymbol{P}_{k-j+1,k-j}  -  \boldsymbol{P}_{k-j+1,k-j+2}=0, \; \text{for} \;  j=1,\ldots,m,
    \end{gathered}
\end{equation}
\begin{equation}
    \label{eq:Gossip1490}
    \begin{gathered}
          \boldsymbol{P}_{k-i+1,k-i+2}=i\cdot \boldsymbol{P}_{k,k+1}-\frac{i-1}{2} + \frac{(i-1)i(i+1)}{6} X % \\
          \quad \text{for} \quad  i=2,\ldots,m
    \end{gathered}
\end{equation}
\begin{equation}
    \label{eq:Gossip1499}
    \begin{gathered}
        \boldsymbol{P}_{k,k+1} = \frac{  3(m+2)  -  m(m+1)(m+2) }  {  6(m+1)  }  X
    \end{gathered}
\end{equation}
\begin{equation}
    \label{eq:Gossip1514}
    \begin{gathered}
            \left(  \left( 2m^3 + 9m^2 + 13m + 6 \right) \cdot X  +  6m^2 + 18m + 15  \right)  \cdot F_{k-m}(X) %\\
            -  6(m+1)^2 \cdot F_{k-m-1}(X)  =  0
    \end{gathered}
\end{equation}
(\ref{eq:Gossip1514}) is the final polynomial for the path topology with even number of vertices.
This is a polynomial of order $k-m+2$ in terms of X.
All roots of this polynomial are negative where $s$ is obtained from the largest root $(X_r)$ of the final polynomial (\ref{eq:Gossip1514}) according to the formula $s=(X_r/2N+1$.
After finding the optimal value of $X$ and correspondingly $s$, the optimal value of probabilities can be obtained by substituting the optimal value of $X$ in formulas (\ref{eq:Gossip1473}), (\ref{eq:Gossip1482}), (\ref{eq:Gossip1490}) and (\ref{eq:Gossip1499}).
The polynomials $F_{k-m}(X)$ and $F_{k-m-1}(X)$ are polynomials of order $k-m-1$ and $k-m-2$, respectively and they can be calculated recursively from equations $F_{1}(X) = 1$, $F_{2}(X) = X+1$ and $F_{i+1}(X) = (X+2) \cdot F_{i}(X) - F_{i-1}(X)$ for $i=2, \ldots, k-m-1$.
Note that $F_{i}(X)$ is a polynomial of order $i-1$ in terms of $X$.
Similar to the symmetric star topology, here for finding the optimal value of $m$, Algorithm \ref{mFindingAlgorithm} should be performed with the final polynomial given in (\ref{eq:Gossip1514}) and the optimal probabilities given in terms of $X$ in equations (\ref{eq:Gossip1473}), (\ref{eq:Gossip1482}), (\ref{eq:Gossip1490}) and (\ref{eq:Gossip1499}).

\subsubsection{Complete Cored Symmetric (CCS) Star Topology}
\label{sec:CCSOptimizationUniform}

In CCS star topology, $n$ path branches of length $k$ (each with $k$ vertices) are connected to each other from one end to form a complete graph at the center of star graph.
A CCS star graph with parameters $k=2$ and $n=5$ is depicted in Figure \ref{fig:GossipFigureSDP}(b).
This topology has $k\cdot n$ vertices in total.
Due to symmetry of graph, the probabilities over edges with same distance from the central core are the same and the transition probabilities over edges of central core are all the same.
Therefore, we have only $2k-1$ variables, namely $\boldsymbol{P}_{1,1}$ and $\boldsymbol{P}_{i,i+1}$, for $i=1,\ldots,k-1$, and $\boldsymbol{P}_{i+1,i}$,for $i=1,\ldots,k-1$.
Note that $\boldsymbol{P}_{1,1}$, is the probability on the edges connecting vertices in the central core of the CCS Star graph and $\boldsymbol{P}_{k,k-1}=1$, since $(i,k)$-th vertex for $i=1,\ldots,n$ is only connected to $(i,k-1)$-th vertex.
Due to the symmetry of CCS star graph towards its central core, we can state that the automorphism group of CCS star graph % $Aut(\mathscr{G})$
is isomorphic to permutation of branches.
We define $m$ as the number of edges (starting from the central core) which their probabilities have not reached the boundaries of feasible region.
Following the same routine performed for symmetric star topology, after solving the equations obtained from complementary slackness conditions we get the following results for the optimal probabilities and second largest eigenvalue $(s)$.
\begin{subequations}
\label{eq:Gossip48}
    \begin{gather}
            1 - (n-1)\cdot \boldsymbol{P}_{1,1} - \boldsymbol{P}_{1,2}  = 0, \label{eq:Gossip48a} \\
            1 - \boldsymbol{P}_{i,i-1} - \boldsymbol{P}_{i,i+1}  = 0,   \quad \text{for} \quad i=2,\ldots, m \label{eq:Gossip48b}
    \end{gather}
\end{subequations}
\begin{equation}
    \label{eq:Gossip49}
    \begin{gathered}
        \boldsymbol{P}_{i+1,i} = -i + \left( \frac{2n\cdot i}{\gamma} + n - 1 \right) \cdot \boldsymbol{P}_{1,1}  %
        +  \left( \frac{i(i+1)}{2\gamma} + \frac{i(i+1)(i-1)}{6} \right) \cdot X, %
        \; \text{for} \; i=1,\ldots,m,
    \end{gathered}
\end{equation}
\begin{equation}
    \label{eq:Gossip50}
    \begin{gathered}
          \boldsymbol{P}_{1,1}  =  \left(  \frac  { \gamma \cdot (m+1) } { 2n\cdot m + \gamma(n-1) }   -
          \left( \frac { 3(m+1)m  +  \gamma\cdot m\cdot(m-1)\cdot(m+1) }  { 12n\cdot m + 6\gamma(n-1) }  \right)\cdot X  \right)
    \end{gathered}
\end{equation}
\begin{equation}
    \label{eq:Gossip51}
    \begin{aligned}
          &\left( \left(12n\cdot\gamma\cdot m^2 + (6(n-1)\cdot\gamma^2 + 12n\cdot\gamma  +12n )\cdot m + 18n\cdot\gamma \right. \right.  \\
          &\qquad\qquad\left. - 6\gamma \right)  + X\cdot\left(  4n\cdot\gamma\cdot m^3 + (3n\cdot(\gamma+1)^2+3n-3\gamma^2 )\cdot m^2  \right.  \\
          &\qquad\qquad\qquad\left. \left.+ (3(n-1)\cdot\gamma^2+(8n-6)\cdot\gamma+6n)\cdot m + 6\gamma\cdot(n-1)\right)\right)    \\
          &\qquad\qquad\qquad\qquad\qquad \cdot  F_{k-m-1}(X) = (6\cdot(1+\gamma\cdot m)\cdot(2n\cdot m+\gamma(n-1) ) )\cdot F_{k-m-2}(X).
    \end{aligned}
\end{equation}
Here $\gamma=\sqrt{2n/(n-1)}$.
The polynomial in (\ref{eq:Gossip51}) is the final polynomial for CCS star topology.
All roots of this polynomial are negative where $s$ is obtained from the largest root $(X_r)$ of the final polynomial (\ref{eq:Gossip51}) according to the formula $s=(X_r/2N)+1$.
After finding the optimal value of $X$ and correspondingly $s$, the optimal value of probabilities can be obtained by substituting the optimal value of $X$ in equations (\ref{eq:Gossip48}), (\ref{eq:Gossip49}), (\ref{eq:Gossip50}).
The polynomials $F_{k-m-1}(X)$ and $F_{k-m-2}(X)$ are polynomials of order $k-m-1$ and $k-m-2$, respectively and they can be calculated recursively  from  equation $F_{0}(X) = 1$, $F_{1}(X) = X+1$ and $F_{i}(X) = (X+2) \cdot F_{i-1}(X) - F_{i-2}(X)$ for $i=2,\ldots,k-m-1$.
Note that $F_{i}(X)$ is a polynomial of order $i$ in terms of $X$.
Similar to the path topology and symmetric star topology, here for finding the optimal value of $m$, Algorithm \ref{mFindingAlgorithm} should be performed with the final polynomial given in (\ref{eq:Gossip51}) and the optimal transition probabilities given in terms of $X$ in equations (\ref{eq:Gossip48}), (\ref{eq:Gossip49}), (\ref{eq:Gossip50}).

In Tables \ref{tab:GossipTableCCS1} and \ref{tab:GossipTableCCS2} we have provided the optimal value of $m$ and $s$ (second largest eigenvalue) for different CCS star topologies.
The results in these tables are presented in order to analyse the optimal value of $m$ and $s$ in terms of topological parameters of CCS star.

\begin{table}
\centering
    \caption {The optimal value of $m$ in terms of number $(n)$ and length $(k)$ of branches for a CCS star topology .}
    \label{tab:GossipTableCCS1}
    \begin{tabular}{|c|c|c|c|c|c|c|} \hline
    {} & {$n=3$} & {$n=4$} & {$n=5$} & {$n=6$} & {$n=7$} & {$n=8$} \\ \hline
    {$k=2$} & {$0$} & {$0$} & {$0$} & {$0$} & {$0$} & {$0$} \\ \hline
    {$k=3$} & {$1$} & {$1$} & {$1$} & {$1$} & {$1$} & {$1$} \\ \hline
    {$k=4$} & {$1$} & {$1$} & {$1$} & {$1$} & {$1$} & {$1$} \\ \hline
    {$k=5$} & {$2$} & {$2$} & {$2$} & {$2$} & {$2$} & {$2$} \\ \hline
    {$k=6$} & {$2$} & {$2$} & {$2$} & {$2$} & {$2$} & {$2$} \\ \hline
    {$k=7$} & {$2$} & {$2$} & {$2$} & {$2$} & {$2$} & {$2$} \\ \hline
    {$k=8$} & {$3$} & {$2$} & {$2$} & {$2$} & {$2$} & {$2$} \\ \hline
    {$k=9$} & {$3$} & {$3$} & {$3$} & {$3$} & {$3$} & {$3$} \\ \hline
    {$k=10$} & {$3$} & {$3$} & {$3$} & {$3$} & {$3$} & {$3$} \\ \hline
    \end{tabular}
\end{table}

\begin{table}
\centering
    \caption {The optimal value of second largest eigenvalue $(s)$ of the gossip operator in terms of number $(n)$ and length $(k)$ of branches for a CCS star topology.}
    \label{tab:GossipTableCCS2}
    \begin{tabular}{|c|c|c|c|c|c|c|} \hline
    {\hspace{-3pt}$k$\hspace{-2pt}}  & {\hspace{-4pt}$n=3$\hspace{-4pt}} & {\hspace{-4pt}$n=4$\hspace{-4pt}} & {\hspace{-4pt}$n=5$\hspace{-4pt}} & {\hspace{-4pt}$n=6$\hspace{-4pt}} & {\hspace{-4pt}$n=7$\hspace{-4pt}} & {\hspace{-4pt}$n=8$\hspace{-4pt}} \\ \hline
    {\hspace{-3pt}$2$\hspace{-2pt}}  & {\hspace{-4pt}$0.95$\hspace{-4pt}} & {\hspace{-4pt}$0.964285$\hspace{-4pt}} & {\hspace{-4pt}$0.972222$\hspace{-4pt}} & {\hspace{-4pt}$0.977272$\hspace{-4pt}} & {\hspace{-4pt}$0.980769$\hspace{-4pt}} & {\hspace{-4pt}$0.983333$\hspace{-4pt}} \\ \hline
    {\hspace{-3pt}$3$\hspace{-2pt}}  & {\hspace{-4pt}$0.982725$\hspace{-4pt}} & {\hspace{-4pt}$0.987533$\hspace{-4pt}} & {\hspace{-4pt}$0.990242$\hspace{-4pt}} & {\hspace{-4pt}$0.991983$\hspace{-4pt}} & {\hspace{-4pt}$0.993196$\hspace{-4pt}} & {\hspace{-4pt}$0.994090$\hspace{-4pt}} \\ \hline
    {\hspace{-3pt}$4$\hspace{-2pt}}  & {\hspace{-4pt}$0.992852$\hspace{-4pt}} & {\hspace{-4pt}$0.994793$\hspace{-4pt}} & {\hspace{-4pt}$0.995903$\hspace{-4pt}} & {\hspace{-4pt}$0.996623$\hspace{-4pt}} & {\hspace{-4pt}$0.997127$\hspace{-4pt}} & {\hspace{-4pt}$0.997500$\hspace{-4pt}} \\ \hline
    {\hspace{-3pt}$5$\hspace{-2pt}}  & {\hspace{-4pt}$0.996396$\hspace{-4pt}} & {\hspace{-4pt}$0.997360$\hspace{-4pt}} & {\hspace{-4pt}$0.997917$\hspace{-4pt}} & {\hspace{-4pt}$0.998279$\hspace{-4pt}} & {\hspace{-4pt}$0.998534$\hspace{-4pt}} & {\hspace{-4pt}$0.998723$\hspace{-4pt}} \\ \hline
    {\hspace{-3pt}$6$\hspace{-2pt}}  & {\hspace{-4pt}$0.997937$\hspace{-4pt}} & {\hspace{-4pt}$0.998483$\hspace{-4pt}} & {\hspace{-4pt}$0.998801$\hspace{-4pt}} & {\hspace{-4pt}$0.999008$\hspace{-4pt}} & {\hspace{-4pt}$0.999154$\hspace{-4pt}} & {\hspace{-4pt}$0.999263$\hspace{-4pt}} \\ \hline
    {\hspace{-3pt}$7$\hspace{-2pt}}  & {\hspace{-4pt}$0.998712$\hspace{-4pt}} & {\hspace{-4pt}$0.999051$\hspace{-4pt}} & {\hspace{-4pt}$0.999248$\hspace{-4pt}} & {\hspace{-4pt}$0.999377$\hspace{-4pt}} & {\hspace{-4pt}$0.999469$\hspace{-4pt}} & {\hspace{-4pt}$0.999536$\hspace{-4pt}} \\ \hline
    {\hspace{-3pt}$8$\hspace{-2pt}}  & {\hspace{-4pt}$0.999143$\hspace{-4pt}} & {\hspace{-4pt}$0.999367$\hspace{-4pt}} & {\hspace{-4pt}$0.999498$\hspace{-4pt}} & {\hspace{-4pt}$0.999584$\hspace{-4pt}} & {\hspace{-4pt}$0.999645$\hspace{-4pt}} & {\hspace{-4pt}$0.999690$\hspace{-4pt}} \\ \hline
    {\hspace{-3pt}$9$\hspace{-2pt}}  & {\hspace{-4pt}$0.999401$\hspace{-4pt}} & {\hspace{-4pt}$0.999557$\hspace{-4pt}} & {\hspace{-4pt}$0.999648$\hspace{-4pt}} & {\hspace{-4pt}$0.999708$\hspace{-4pt}} & {\hspace{-4pt}$0.999751$\hspace{-4pt}} & {\hspace{-4pt}$0.999783$\hspace{-4pt}} \\ \hline
    {\hspace{-3pt}$10$\hspace{-2pt}} & {\hspace{-4pt}$0.999566$\hspace{-4pt}} & {\hspace{-4pt}$0.999678$\hspace{-4pt}} & {\hspace{-4pt}$0.999744$\hspace{-4pt}} & {\hspace{-4pt}$0.999788$\hspace{-4pt}} & {\hspace{-4pt}$0.999819$\hspace{-4pt}} & {\hspace{-4pt}$0.999842$\hspace{-4pt}} \\ \hline
    \end{tabular}
\end{table}

Similar to Symmetric star topology, in CCS star topology, increasing either the length or the number of branches degrades the convergence rate of the randomized gossip algorithm.

\subsubsection{Two Coupled Complete Graphs}
\label{sec:TwoCoupledOptimizationUniform}

In this topology, two complete graphs each with $N_1 + N_2$ and $N_2 + N_3$ vertices respectively, share $N_2$ vertices.
In figure \ref{fig:GossipFigureSDP}(c) two coupled complete graphs with parameters $N_1 = 4$, $N_2 = 2$ and $N_3 = 3$ is depicted.
This topology has $N = N_1 + N_2 + N_3$ vertices.
Due to the symmetry of the complete graphs, the transition probabilities can be divided into seven groups.
$\boldsymbol{P}_{-1,-1}$ is the transition probability on edges connecting the $N_1$ vertices on the top complete graph to each other.
$\boldsymbol{P}_{-1,0}$ is the transition probability on the edges connecting the $N_1$ vertices on the top complete graph to the $N_2$ vertices in the middle and $\boldsymbol{P}_{0,-1}$ is the transition probability on the same edges as in $\boldsymbol{P}_{-1,0}$ but with reverse direction.
$\boldsymbol{P}_{0,0}$ is the probability on edges connecting the $N_2$ vertices in the middle to each other.
Similarly the transition probabilities $\boldsymbol{P}_{0,1}$, $\boldsymbol{P}_{1,0}$ and $\boldsymbol{P}_{1,1}$ are defined for the transition probabilities on the edges of the complete graph on the down side of the topology.

For the symmetric case where $N_1 = N_3$ the optimal transition probabilities and $\lambda_2(\overline{\boldsymbol{W}})$ are obtained for two following categories.
For $N_2 > 2N_1$, the optimal probabilities are $\boldsymbol{P}_{-1,-1}  =  \boldsymbol{P}_{1,1} = 0$, $\boldsymbol{P}_{1,0} = \boldsymbol{P}_{-1,0} = 1/N_2$, $\boldsymbol{P}_{0,1} = \boldsymbol{P}_{0,-1} = (  2N_2^2 - (N_2 - 1)(N_2 - 2N_1)  )  /  (  N_2\left( 4 N_1 N_2 + (N_2 - 1) (N_2 - 2N_1) \right)  )$, $\boldsymbol{P}_{0,0}  =  \left( (2N_1 + N_2)(N_2 - 2N_1) \right) / \left( N_2 \left( 4 N_1 N_2  +  (N_2 - 1)(N_2 - 2N_1) \right) \right)$,
and the optimal value of the second largest eigenvalue of the gossip operator $(\lambda_2(\overline{\boldsymbol{W}}))$ is %as below,
$\lambda_2(\overline{\boldsymbol{W}}) = (  4 N_1 N_2 + (N_2 - 1)(N_2 - 2 N_1) - N_2  )  /   (  4 N_1 N_2 + (N_2 - 1) (N_2 - 2N_1)  )$.

For $N_2 \leq 2N_1$, the transition probabilities are % as below,
$\boldsymbol{P}_{-1,-1}  =  \boldsymbol{P}_{1,1} = \boldsymbol{P}_{0,0} = 0$, $\boldsymbol{P}_{1,0} = \boldsymbol{P}_{-1,0} = 1/N_2$, $\boldsymbol{P}_{0,1} = \boldsymbol{P}_{0,-1} = 1/(2N_1)$,
and the optimal value of the second largest eigenvalue of the gossip operator $(\lambda_2(\overline{\boldsymbol{W}}))$ is $\lambda_2(\overline{\boldsymbol{W}}) = (  4 N_1 - 1  ) /   (  4 N_1  )$.
It is apparent that for the last symmetric case where $N_2 \leq 2N_1$ the whole topology reduces to a 3-partite graph.

\subsection{Optimization with Non-Uniform Clock Distribution}
\label{sec:OptimizationSectionNonUniform}

In this section we address the optimization of the FCGA problem (\ref{eq:ClassicalGossipSDP}) with the assumption that the clock distribution of vertices is non-uniform (i.e. $\boldsymbol{P}_i \neq 1/N$).

\subsubsection{Symmetric Star Topology}
The symmetric star topology with parameters $n$ and $k$ is introduced in subsection \ref{sec:SymmetricStar}.
As explained in subsection \ref{sec:SymmetricStar}, all edges with the same distance from the central vertex have the same optimal transition probability.
For non-uniform selection of vertices, the optimal value of the second largest eigenvalue of the gossip operator is  $(\lambda_2(\overline{\boldsymbol{W}}))  =  1-  3/(nk(k+1)(2k+1))  $
and the optimal transition probabilities are % as below, % $q_{j-1,j}$  are as below,
$\boldsymbol{P}_{0,1} = 1/n$, $\boldsymbol{P}_{1,0} = 1$, $\boldsymbol{P}_{j-1,j}= 0$, $\boldsymbol{P}_{j,j-1}= 1$ for $j=2, \ldots, k$.
The optimal clock distribution probabilities of vertices are
$\boldsymbol{P}_0 = 0$, $\boldsymbol{P}_{j}  =  \left(3(k+j)(k-j+1)\right) / \left(nk(k+1)(2k+1)\right)$  for  $j=1, \ldots, k$.
For nonzero $\boldsymbol{P}_0 < 3/(2k+1)$, the optimal clock distribution probability $\boldsymbol{P}_1$ is $\boldsymbol{P}_1  =  3/(n(2k+1)) - \boldsymbol{P}_0 / n$ and the rest of the optimal clock distribution probabilities are same as the case of $\boldsymbol{P}_0 = 0$.
A special case of the symmetric star topology is the case of $n=2$, where this topology reduces to a path topology with odd number of vertices (i.e. $N=2k+1$).
For this case, the optimal value of the second largest eigenvalue of the gossip operator is % as below,
$\lambda_2 ( \overline{\boldsymbol{W}} ) = 1 -  6/(N(N-1)(N+1))$ and the optimal transition probabilities are $\boldsymbol{P}_{0,1}= 1/n$, $\boldsymbol{P}_{1,0} = 1$, $\boldsymbol{P}_{j-1,j}= 0$ and $\boldsymbol{P}_{j,j-1}= 1$ for $j=2, \ldots, k$.
The optimal clock distribution probabilities are $\boldsymbol{P}_0=0$ and $\boldsymbol{P}_j=  (3(N+2j-1)(N-2j+1))/(N(N-1)(N+1))$ for $j=1, \ldots, k$.
For nonzero $\boldsymbol{P}_0 < 3/N$, the optimal clock distribution probability $\boldsymbol{P}_1$ is $\boldsymbol{P}_1  =  3/(n(2k+1))-\boldsymbol{P}_0 / 2 $ and the rest of the optimal clock distribution probabilities are same as the case of $\boldsymbol{P}_0 = 0$.

\subsubsection{Complete-Cored Symmetric Star Topology}

The Complete-Cored Symmetric (CCS) star topology with parameters $(n,k)$ is introduced in subsection \ref{sec:CCSOptimizationUniform}.
As explained in subsection \ref{sec:SymmetricStar}, all edges (vertices) with the same distance from the central core have the same optimal transition (clock distribution) probabilities.

For non-uniform clock distribution, the optimal transition probabilities are $\boldsymbol{P}_{0,0} = 1/(n-1)$, $\boldsymbol{P}_{j,j-1}=1$ and $\boldsymbol{P}_{j-1,j} = 0$ for $j=1,\ldots, k$.
The optimal clock distribution probabilities are $P_0  =  3 ( 2n - 2 + k\sqrt{2n(n-1)} ) 	/ 	(	2n ( 3n - 3 + 3k\sqrt{2n(n-1)}  +  2nk^2  +  nk )	)$ and $P_j  =     3  (  \sqrt{2n(n-1)} ( k-j+1 ) + n(k-j+1)(k+j)  )	/    (	3n(k+1)( n-1 + k\sqrt{2n(n-1)} )  +  n^{2}k(k+1)(2k+1)	)$ for $j = 1, \ldots, k$.
The optimal value of the second largest eigenvalue of the gossip operator is $(\lambda_2(\overline{\boldsymbol{W}}))  =   1- 3  / (3(n-1)(k+1)+3\sqrt{2n(n-1)}k(k+1)+nk(k+1)(2k+1))$.
A special case of the CCS star topology is the path topology with even number of vertices which is obtained for $n=2$.
For the path topology with $2(k+1)$ vertices, the optimal clock distribution probabilities are $\boldsymbol{P}_0  =  3 (k+1) 	/ 		(2(2k+3)(2k+1))	$ and $\boldsymbol{P}_j  =  3  (  (k+1)^2 - j^2  )	/     (	(k+1)(2k+1)(2k+3)	)$ for $j=1, \ldots, k$ and the second largest eigenvalue of the gossip operator is $\lambda_2(\overline{\boldsymbol{W}}) = 1- 3 / ((k+1)(2k+1)(2k+3))$.

\subsubsection{CCS Star with two types of branches}

The Complete-Cored Symmetric (CCS) star with two types of branches is identified with parameters $(n, k_1, k_2)$.
This topology is a CCS star topology where two types of tails (each with $k_1$ and $k_2$ edges) are connected to each vertex in the complete core.
A CCS star graph  with two types of branches with parameters $n=5$, $k_1=2$ and $k_2 = 1$ is depicted in figure \ref{fig:GossipFigureSDP}(e).

For non-uniform clock distribution, the optimal transition probabilities are $\boldsymbol{P}_{0,0}=1/(n-1)$, $\boldsymbol{P}_{j,j-1}=1$, $\boldsymbol{P}_{j-1,j}=0$ for $j = 1, \ldots, k_1$, and $P_{j+1,j}=0$, $P_{j,j+1}=1$ for $j = -1, \ldots, -k_2$.
The optimal second smallest largest eigenvalue of the gossip operator is $ \lambda_2 (\overline{\boldsymbol{W}})  =  1- 3  /  (    3(n-1)(k_1 + k_2 + 1) + 3\sqrt{2n(n-1)}D_1  +  n D_2    ) $ with $D_1  =   k_1(k_1+1) + k_2(k_2+1)$ and $D_2  =   k_1 ( k_1 + 1 ) ( 2k_1 + 1 )  +   k_2 ( k_2 + 1 ) ( 2k_2 + 1 )$.
The optimal transition probabilities are $\boldsymbol{P}_0 = (1-\lambda_2(\overline{\boldsymbol{W}})) \times ( 2(n-1)(k_1+k_2+1) + \sqrt{2n(n-1)}(k_1(k_1+1)+k_2(k_2+1)) ) /  2n$, $\boldsymbol{P}_j  =  (1-\lambda_2(\overline{\boldsymbol{W}})) \times  ( \sqrt{2n(n-1)}(k_1+j+1) + n(k_1+j+1)(k_1-j ) ) / n $ for $j=-k_1,\ldots,-1$ and $\boldsymbol{P}_j  =  (1-\lambda_2(\overline{\boldsymbol{W}})) \times (\sqrt{2n(n-1)}(k_2-j+1) + n(k_2-j+1)(k_2+j )) / n$ for $j=1,\cdots,k_2$.

\subsubsection{Palm Topology}
\label{sec:PalmGraph}
A palm topology with parameters $(n,k)$ consists of a path graph with $k$ vertices connected to the central vertex of a star graph with $n$ branches as shown in figure \ref{fig:GossipFigureSDP}(d) for parameters $n=4$ and $k=2$.
This topology has $n+k+1$ vertices and $n+k$ edges.
Due to the symmetry of the topology, all edges connected to the central vertex of the star graph have the same transition probabilities (denoted by $\boldsymbol{P}_{0,-l}$ and $\boldsymbol{P}_{-1,0}$) except the edge connecting the path graph to the central vertex and the vertices on the star graph (other than the central vertex) have the same clock distribution probability.
By assigning the transition probabilities as in $\boldsymbol{P}_{j,j-1}=1$, $\boldsymbol{P}_{j-1,j}=0$ for $j=1, \ldots, k$ and $\boldsymbol{P}_{-j,0}=1$, $\boldsymbol{P}_{0,-j}=0$ for $j=1, \ldots, n$,
the optimal answer varies, depending on the values of the parameters $n$ and $k$.
For $2n > k(k+1)$ the optimal value of the second largest eigenvalue of the gossip operator is $\lambda_2(\overline{\boldsymbol{W}})  =   1-  3 /  (  6n + k(k+1)(2k+1)  )$, and the optimal clock distribution probabilities are $\boldsymbol{P}_0  = 0$, $\boldsymbol{P}_{-l}= 6 /  (  6n + k(k+1)(2k+1))$, for $l=1, \ldots, n$ and $\boldsymbol{P}_j  =    (   3(k-j+1)( k+j)    ) /  ( 6n+k(k+1)(2k+1)     )$,  for $j=1, \ldots, k$.
For $2n \leq k(k+1)$ the optimal value of the second largest eigenvalue of the gossip operator is $\lambda_2(\overline{\boldsymbol{W}}) = 1- ( 6(n+k+1) )  /  (  (k+1)(k+2)(6n + k(k+4n+1))  )$,
and the optimal clock distribution probabilities are $\boldsymbol{P}_0  = 0$, $\boldsymbol{P}_{l}= (1-\lambda_2(\overline{\boldsymbol{W}})) \cdot ( (k+1)(k+2)  /   (n+k+1) )$ for $l=-1, \ldots, -n$ and $\boldsymbol{P}_j  =  (1-\lambda_2(\overline{\boldsymbol{W}})) \cdot ( ( k-j+1 ) ( n(k+j+2) +(k+1)j ) )  /   (n+k+1) $, for $j=1,\ldots, k$.

\subsubsection{Lollipop Topology}

This topology is obtained by connecting a path graph (with $k$ vertices) to one of the vertices in a complete graph with $n+1$ vertices.
By bridging vertex, we refer to the vertex in complete graph that is connected to the path graph.
Considering the symmetry of the complete graph the transitive probabilities in the complete graph can be categorized into two groups.
The first group is those connecting the vertices in the complete graph other than the bridging vertex.
We denote the transitive probability on these edges by $\boldsymbol{P}_{-1,-1}$.
The second group is the edges from the bridging vertex to other vertices in the complete graph and vice versa, where the transitive probability over these edges are denoted by $\boldsymbol{P}_{0,-1}$ and $\boldsymbol{P}_{-1,0}$, respectively.
The transitive probabilities on the edges of the tail are denoted by $\boldsymbol{P}_{j,j-1}$ and $\boldsymbol{P}_{j-1,j}$ for $j=1 \ldots, k$.
The Lollipop topology is depicted in figure \ref{fig:GossipFigureSDP}(f) along with the probabilities assigned to the edges.
For the case that $k(k+1) > \sqrt{2n(n+1)}$ the optimal value of the transition probability $\boldsymbol{P}_{-1,-1}$ is zero and the Lollipop topology reduces to Palm topology where the optimal answer for this topology is provided in section \ref{sec:PalmGraph}.
For the case of $k(k+1) \leq \sqrt{2n(n+1)}$, by selecting the transition probabilities as $\boldsymbol{P}_{j,j-1}=1$,  $\boldsymbol{P}_{j-1,j}=0$ for $j = 1, \ldots, k$ and $\boldsymbol{P}_{-1,0}=0$, $\boldsymbol{P}_{0,-1}=1/n$, $\boldsymbol{P}_{-1,-1}=1/(n-1)$,
the optimal value of the second largest eigenvalue of the gossip operator is obtained as $\lambda_2(\overline{\boldsymbol{W}})  =  1- 6(n+k+1)/A$ where $A$ is $A = 6(n-1)(n+k+1)  +  (k+1) ( 6k \sqrt{2n(n+1)} + (n+1)(6 + k(k+2)) + k^2(3n+k+2))   $,
and the optimal transition clock distribution probabilities are $\boldsymbol{P}_0  = (1-\lambda_2(\overline{\boldsymbol{W}})) n(k+1) ( 2(n+1) + k\sqrt{2n(n+1)} )/ ((n+k+1)(n+1))$, $\boldsymbol{P}_{l}  =  ((n-1)(1-\lambda_2(\overline{\boldsymbol{W}})-\boldsymbol{P}_0/(2n)))/n$, for $l=-1, \cdots, -n$ and $\boldsymbol{P}_j  =  (1-\lambda_2(\overline{\boldsymbol{W}}))     (k-j+1)            \left(  \sqrt{2n(n+1)}  +  j(k+n+1)+ nk  \right) / (n+k+1)$  for $j=1,\ldots,k$.

\subsection{Detailed Balance}

For every underlying topology, there is a set of clock distribution and transition probabilities for the global optimal solution of the gossip algorithm.
A straightforward method to obtain these probabilities is based on the SDP formulation of the continuous time consensus algorithm \cite{SaberQConsensusContinuous} equivalent to that of that of the gossip algorithm.
The equivalent SDP formulation of gossip algorithm can be obtained from that of the continuous time consensus algorithm by setting the upper limit on the total amount of weights $(D)$ to $1/2$ and defining the clock distribution and transition probabilities $(\boldsymbol{P}_{i}, \boldsymbol{P}_{i,j})$ in terms of the weights of the continuous time consensus algorithm
as below,
\begin{subequations}
    \label{eq:DetailedBalanced}
    \begin{gather}
        \boldsymbol{P}_{i}  =  \sum_{j \in N(i)} { w_{i,j} },         \\
        \boldsymbol{P}_{i,j}  =  \frac  {   w_{i,j}   }  {    \boldsymbol{P}_{i}    },
     \end{gather}
\end{subequations}
where $N(i)$ is the set of neighbours of $i$-th vertex in the underlying graph of the network.
It is obvious that $\sum_{i}{\boldsymbol{P}_{i}} = \sum_{i}{\sum_{j \in N(i)}{\boldsymbol{w}_{i,j}}}  =  2 \times \left( \sum_{\{i,j\}\in \mathcal{E}} {\boldsymbol{w}_{i,j}} \right)  =  2 \times D  =  1$.
Also for variable $q_{i,j}$ we have $q_{i,j} = ( \boldsymbol{P}_{i}\boldsymbol{P}_{i,j} + \boldsymbol{P}_{j}\boldsymbol{P}_{j,i} )/2  =  \boldsymbol{w}_{i,j}$ and since $\boldsymbol{w}_{i,j} = \boldsymbol{w}_{j,i}$ then $\boldsymbol{q}_{i,j} = \boldsymbol{q}_{j,i}$.
By choosing the clock distribution and transition probabilities as in (\ref{eq:DetailedBalanced}), the following can be concluded for the second largest eigenvalue of the gossip operator $(\overline{\boldsymbol{W}})$ and the second smallest eigenvalue of the Laplacian matrix in the continuous-time consensus problem,
\begin{equation}
    \nonumber
    \begin{aligned}
        \lambda_{2}(\overline{\boldsymbol{W}})  =  1 - \lambda_{2}(\boldsymbol{L}).
    \end{aligned}
\end{equation}
For the probabilities (\ref{eq:DetailedBalanced}), it can be shown that $\boldsymbol{P}_{i} \cdot \boldsymbol{P}_{i,j} = \boldsymbol{P}_{j} \cdot \boldsymbol{P}_{j,i}$, therefore, the selection of the probabilities in (\ref{eq:DetailedBalanced}) is said to have the detailed balance property \cite{DetailedBalanceRef}.
This is due to the fact that
for all pairs of states, $\boldsymbol{P}_{i} \boldsymbol{P}_{i,j}$ (the rate at which transitions occur from
state $i$ to state $j$) balances $\boldsymbol{P}_{j} \boldsymbol{P}_{j,i}$ (the rate at which transitions occur from $j$ to state $i$).

Detailed balance is intimately related to time reversibility in Markov chain.
Consider a continuous-time Markov chain $\{X(t)\}$ with unique equilibrium distribution $\{P_i\}$ and transition probabilities $\{P_{ij}\}$.
The continuous-time Markov chain $\{X(t)\}$ is said to satisfy detailed balance property if the following holds,
\begin{equation}
    \label{eq:3151}
    \begin{aligned}
        	P_i P_{ij} = P_j P_{ji}  \quad \text{for all} \quad i,j \quad \text{with} \quad i \neq j.	
    \end{aligned}
\end{equation}
In other words, the long-run average number of transitions from state k to state $i$ per time unit is equal to the long-run average number of transitions from state $i$ to state $j$ per time unit for all $i \neq j$.
To explain the time reversibility, consider the stationary Markov chain $\{X(t)\}$, i.e. $P\{X(t) = i\} = P_{i}$, for all $t \geq 0$.
It is shown \cite{PoissonBook} that the condition (\ref{eq:3151}) is satisfied if and only if the stationary Markov chain $\{X(t)\}$ has the property that for all $n \geq 1$ and all $u > 0$,
$ (X(u_1), \ldots, X(u_n)) $ is distributed as $ (X(u - u_1), \ldots, X(u - u_n)) $ for all $0 < u_1 < \cdots < u_n < u$.
In other words, the process reversed in time has the same probabilistic structure as the original process when the process has reached statistical equilibrium.
A Markov process with this property is said to be time reversible.

\section{Conclusion}
\label{sec:Conclusion}

We have optimized the convergence rate of the gossip algorithm for both classical and quantum networks.
In contrast to previous models of the gossip algorithm in literature, in this paper we have proposed a novel model of the gossip algorithm with non-uniform clock distribution.
This is done to propose and optimize a model for gossip algorithm that can achieve the convergence rate of the optimal continuous-time consensus algorithm over the same network topology.
By modeling the clock tick model of the gossip algorithm as a Poisson process, we have explained how non-uniform clock distribution can be achieved by modifying the rate of the Poison process according to the clock distribution probabilities.
Deriving the minimization problem for optimizing the asymptotic convergence rate of the proposed gossip algorithm, we have analytically addressed its corresponding semidefinite programming formulation for different topologies and obtained closed-form formulas for the probabilities and the convergence rate of the gossip algorithm for both cases of uniform and non-uniform clock distributions.
Based on the results provided, it can be concluded that the optimal results obtained for uniform clock distribution are suboptimal compared to those of the non-uniform one.
Since the number of variables in the semidefinite programming formulation of the problem is more than the number of equations, the optimal answer for non-uniform clock distribution is not unique.
In other words, there is more than one set of transition and clock distribution probabilities that can achieve the fastest convergence rate.
Based on the optimal results of continuous-time consensus algorithm \cite{SaberQConsensusContinuous}, a method for achieving one of the optimal results for the case of non-uniform clock distribution is proposed.
The proposed method ensures that the transition and clock distribution probabilities satisfy detailed balance property.

In the case of the gossip algorithm for quantum networks, by expanding the density matrix in terms of the generalized Gell-Mann matrices, we transform the evolution equation of the quantum gossip algorithm to the state update equation of the classical gossip algorithm.
Defining the gossip operator for quantum gossip algorithm, we have formulated the original optimization problem as a convex optimization problem, which can be solved using semidefinite and linear programming.

\end{document}